\documentclass[11.5pt]{article}
\usepackage{amssymb}
\usepackage[pdftex]{graphicx}
\usepackage{overpic}
\usepackage{bm}
\usepackage{physics}
\usepackage[paper=letterpaper,margin=0.90in]{geometry}
\usepackage{amsfonts,amsthm,amsmath,amssymb,upgreek}
\usepackage{units}
\usepackage{here}

\usepackage{amsfonts,amsthm,amsmath,amssymb,upgreek}
\usepackage{graphicx}
\usepackage{units}
\usepackage{float}
\usepackage [english]{babel}
\usepackage [autostyle, english = american]{csquotes}

\numberwithin{equation}{section}

\def\hat{\widehat}
\newcommand{\be}{\begin{equation}}
\newcommand{\ee}{\end{equation}}

\title{\Large  What happens due to the baby universe effect in JT gravity?\\
- \large{Analysis of correlation functions and ERB length at late time using three approaches} -}
\author{Masayoshi Sato\footnote{ed23003@nda.ac.jp}}
\date{\small Department of Applied Physics, National Defense Academy,\\ 1-10-20 Hashirimizu, Yokosuka-city, Kanagawa 239-8686 Japan}

\begin{document}
\setlength{\abovedisplayskip}{-1pt}
\setlength{\belowdisplayskip}{-1pt}

\maketitle
\begin{abstract}
We analyze the correlation function in JT gravity using three approaches: by summing over all geodesics connecting boundary operators, integrating over the region of moduli space determined by the ``no-shortcut condition'' introduced in \cite{Firewall-1}, and using the formula for the universal spectral density correlation in the $\tau$-scaling limit. We find that the behaviors of the three results coincide at late times: they all exhibit a ``ramp'' instead of permanent decay. Using the third approach we also confirm that the ``plateau'' appears after $T_H=2\pi e^{S_0}\hat{\rho}_0(E)$. Overall, our results are consistent with the SFF analysis.

We also calculate the ERB length $\langle \ell(T) \rangle$ using the three approaches and find that the results are in good agreement with each other. We also find that the $\langle \ell(T) \rangle$ grows as a cubic function in $T$ due to the contribution from geometry including one observable baby universe, and converges to a constant after $T=T_H$. For the geometry with one baby universe, we compute the size $\langle b(T) \rangle$ of the baby universe and find that it is of the same order as $\langle \ell(T) \rangle$. This result is consistent with the baby universe emission mechanism claimed in \cite{Late}.
\end{abstract}
\tableofcontents 
\section{Introduction} \label{Intro}
When an eternal black hole in AdS spacetime is perturbed, the two-point function of the quantum field observed outside the black hole changes over time \cite{Info-Malda2001}. In semi-classical theory, the two-point function is expected to decay indefinitely over time \cite{Info-Horowitz1999}. However, according to the AdS/CFT correspondence, this decay does not persist permanently. At late times, it has a non-zero value, but since it is extremely small relative to the entropy of the system, it cannot be observed by perturbation theory and it is difficult to describe its exact behavior \cite{Info-Malda2001,Info-Goh2002,Info-Dyson2002,Info-Barbon2003}.
 
On the other hand, in ensemble-averaged theories such as the SYK model \cite{Sach1992,Kitaev2015,Polchinski2016,Malda2016,Kitaev2017,Cotler2016,Saad2018} one can consider the average behavior of the two-point correlation. If the Hamiltonian of this ensemble theory obeys the statistics of a random matrix, the spectral form factor (SFF: the ensemble average of the product of partition functions $\langle Z(\beta+iT)Z(\beta-iT)\rangle$) shows an interesting behavior as a function of time $T$.  When it reaches at late-time, it stops decaying and undergoes a linear growth called the ``ramp" \cite{Cotler2016}, and eventually reaches a ``plateau" and converges to a constant.

The aforementioned phenomena can be studied further by analyzing the two-point correlation function \cite{Malda2016-2,Yang,Gross2017,Lam2018,Mertens2017,corre-1,corre-2,corre-3,Bulycheva2019,Iliesiu2019} in the Jackiw-Teitelboim (JT) gravity coupled to matter fields \cite{JT1,JT2,Almheiri2014,Malda2016-2}. It has been confirmed that this correlation function exhibits a non-decaying behavior at late times due to a quantum gravity effect. This behavior is, in fact, attributed to topological changes caused by Euclidean wormholes \cite{Hawking1987,Giddings-1,Giddings-2,Cole-baby-1,Cole-baby-2,Klebanov1988,Malda2004,ArkaniHamed2007}. This Euclidean wormhole is produced by a tunneling process in which a closed baby universe is emitted or absorbed from its parent AdS universe.

Regarding the actual computation of correlators in JT gravity, there seem to be several natural approaches. On one hand, in \cite{Late,Volume} the correlator was defined as a sum over all (non self-intersecting) geodesics connecting the two operators. On the other hand, in 
\cite{Firewall-1} the handle-disk contribution to the JT partition function was evaluated by first selecting a specific geodesic connecting the two boundary points and regarding it as the spatial slice. One then integrates over a certain region of the moduli space determined by the ``no-shortcut condition'' which ensures that the chosen geodesic is the shortest. By suitably modifying their computation, one may define the correlator as the expectation value of $e^{-\Delta \ell}$ with $\ell$ the length of the shortest geodesic. Though this is a slightly non-standard definition, it has a finite $\Delta \to 0$ limit and we expect it to be more suited to studying observables such as $\langle \ell \rangle$. And there is yet another independent approach which uses the formula for universal spectral density correlation \cite{SSS} of random matrices in the $\tau$-scaling limit \cite{OS}. In contrast to the first two perturbative approaches, this method allows us to work out the all-genus result at once, though each term in the perturbation series gets somewhat simplified. It would be interesting to clarify to what extent the results obtained from these three approaches agree with each other.

In this paper we analyze the correlator using the methods explained above. The analysis at late times near $T_H=2\pi e^{S_0}\hat{\rho}_0(E)$ shows good agreement between the three approaches. All three results for the correlation function indicate a ``ramp'' and no permanent decay. Also, the calculations using the third method reveal the ``plateau'' where the correlation function converges to a constant at very late time. This behavior is hard to capture using the other (perturbative) approaches, for which this paper focuses only on the first subleading term. Overall, our results are consistent with the SFF analysis.

Also, in \cite{Late} it was claimed that after an Einstein-Rosen Bridge (ERB) has grown for a long time, a part of it is cut off and emitted as a baby universe, and  its size $\langle b(T) \rangle$ is very close to the length of the ERB at the time of emission. Thus, we calculate $\langle \ell(T)\rangle$ using the three methods. We find that the $\langle \ell(T) \rangle$ grows as  a cubic function in $T$ due to the contribution from geometry including one observable baby universe, and converges to a constant after $T=T_H$. This result provides valuable information for elucidating the properties of ERB. And finally, we compute the size $b(T)$ of the baby universe and confirm that it is of the same order as $\langle \ell(T) \rangle$. This result is consistent with the baby universe emission mechanism claimed in \cite{Late}.

Let's outline the structure of this paper. In Section 2 we review the JT gravity in Lorentzian form and summarize the formula needed for computing correlators. In Section 3 we introduce the method for calculating two-point correlation functions according to \cite{Late,Volume}. For the purpose of studying the baby universe, we regard the geodesic connecting the two boundary operators as the spatial slice (equal-time slice). We regard there is an observable baby universe when this slice ``passes through'' the handle, and distinguish different contributions to the two-point correlation functions between those which contain baby universes and those which do not. In Section 4 we calculate the correlation function using three different methods and analyze its behavior. We show that the results from the three methods agree with each other, and  by combining these three results we are able to recover the ramp and plateau behavior at late times. In Section 5 we calculate the expectation value of the ERB length and the size of baby universe, and show that the two have the same order of magnitude at late times. Finally, in section 6, we discuss the physical insights gained from comparing the three different methods and some future directions. In the Appendices we collect some details of the derivations of our results as well as a review of analysis of SFF in JT gravity \cite{SFF-APP-4}.
\section{Review of JT gravity} \label{Review}
\subsection{\normalsize Genus expansion of the partition function}
In this section we review JT gravity, with a focus on the way the basic observables are expressed in terms of genus expansion. The action of a general two-dimensional dilaton gravity takes the following form and is characterized by the dilaton potential $U(\phi)$:\\
\begin{align}
I_{\text{2D}}= -S_0\chi(\mathcal{M}) - \Big[\ \frac{1}{2}\int_{\mathcal{M}}d^2 x\sqrt{g}(\phi R-U(\phi))+\int_{\mathcal{\partial M}}dx\sqrt{h}\phi K \ \Big], \label{1-1} \\ \nonumber
\end{align}
where we have set $8\pi G_N=1$. The first term represents the topological term which arises from the Einstein-Hilbert term. JT gravity is defined by setting the dilaton potential as $U(\phi) = -2\phi$, corresponding to spacetimes with locally AdS$_2$ metric $ds^{2}=\frac{dT^2+dZ^2}{Z^2}$. In this case, the JT action on $\mathcal{M}$ is expressed as follows \cite{JT1,JT2,Muta:1992xw,Nojiri:2024ycf}: \\
\begin{align}
I_{\text{JT}}[g,\phi]=-\frac{1}{2}\int_{\mathcal{M}}d^2 x \sqrt{g}\ \phi(R+2)-\int_{\partial \mathcal{M}}dx \sqrt{h} \ \phi(K-1). \label{1-2} \\ \nonumber 
\end{align}
Then, we regularize the integral over the infinite volume by shifting the holographic boundary slightly inward from $Z = 0$ to $Z =\epsilon \ (\epsilon \to 0)$ and imposing the boundary conditions $ds^{2}|_{\partial \mathcal{M}}=d\tau^2/\epsilon^2$ and $\phi_{\partial \mathcal{M}}=1/\epsilon$ on the metric and dilaton, respectively. The boundary term then turns into the Schwarzian action, where the coordinate $\tau \sim \tau+\beta$ is a rescaled proper length coordinate along the boundary. Note that we introduced $-\int_{\partial \mathcal{M}}dx \sqrt{h}\phi$ as a counter term in (\ref{1-2}).

The partition function is given as a sum over different topologies characterized by the genus $g$ and the number of boundaries $n$: \\
\begin{align}
Z(\beta_1,\dots,\beta_n)_{\text{conn}} = \sum_{g=0}^{\infty }\ e^{S_{0}(2-2g-n)}Z_{g,n}(\beta_1,\dots,\beta_n). \label{1-3} 
\end{align}
\\
The partition function $Z_{g,n}(\beta_1, \dots, \beta_n)$ for genus $g$ with $n$ boundaries is evaluated as the path integral over the dilaton and the metric: \\
\begin{align}
Z_{g,n}(\beta_1,\dots \beta_n)=\int \frac{\mathcal{D}\phi \, \mathcal{D}g_{\mu\nu}}{\text{Vol(Diff)}} \,e^{-I_{\text{JT}}[g,\phi]}, \label{1-4}
\end{align}
\\
where $\frac{\mathcal{D}g_{\mu\nu}}{\text{Vol(Diff)}}$ represents the integration measure that takes into account the redundancy associated with the diffeomorphism group. In this integration, one first integrates out the dilaton, which generates the constraint $\delta(R+2)$. This delta function fixes the geometry to locally AdS$_2$ spacetime. One is then left with the integration over the fluctuations of the AdS$_2$ boundary shapes (boundary wiggles) and the moduli space of the bulk geometry, as follows:\\
\begin{align}
Z_{g,n}(\beta_1,\dots,\beta_n) = \int \mathcal{D}(\text{bulk moduli})\int \mathcal{D}(\text{boundary wiggles})\ e^{\int_{\partial \mathcal{M}}\sqrt{h}\phi(K-1)}. \label{1-5} \\ \nonumber
\end{align}
By evaluating the integral over the boundary shapes explicitly, the disk and trumpet partition functions were obtained in \cite{Disk-par,SSS}. Then, by connecting $n$ trumpet partition functions to the Weil-Petersson volume $V_{g,n}$ of the moduli space, the partition function corresponding to genus $g$ with $n$ boundaries can be expressed as follows:\\
\begin{align}
& Z_{0,1}(\beta)=Z_{D}(\beta)=\ \frac{e^{2\pi^2/\beta}}{\sqrt{2\pi}\beta^{3/2}}, \quad Z_{\text{Tr}}(\beta,b)=\ \frac{e^{-b^2/2\beta}}{\sqrt{2\pi \beta}}, \label{1-6} \\ 
& Z_{g,n}(\beta_1,\dots,\beta_n) =\  \int_{0}^{\infty}\prod_{i=1}^{n}b_i db_i V_{g,n}(b_1,\dots,b_n)Z_{\text{Tr}}(\beta_i,b_i). \label{1-7}  \\ \nonumber
\end{align} 
In particular, the one-point and  connected two-point functions are given by \cite{SSS}:\\
\begin{align}
\big \langle Z(\beta)\big \rangle &= \ e^{S_0}\ Z_{D}(\beta)+\sum_{g=1}^{\infty}e^{(1-2g)S_0}\int_{0}^{\infty}db\,b\ V_{g,1}(b) Z_{\text{Tr}}(\beta,b), \label{1-8} \\
\big \langle Z(\beta_1) Z(\beta_2)\big\rangle_{\text{conn}} & = \ Z_{0,2}(\beta_1,\beta_2)+\sum_{g=1}^{\infty}e^{-2g S_0}Z_{g,2}(\beta_1,\beta_2) \nonumber \\
&= \ \int^{\infty}_{0} bdb \ Z_{\text{Tr}}(\beta_1,b)Z_{\text{Tr}}(\beta_2,b) + \sum_{g=1}^{\infty}e^{-2g S_0} \int_{0}^{\infty}\prod_{i=1}^{2}b_i db_i V_{g,2}(b_1,b_2)Z_{\text{Tr}}(\beta_i,b_i) .\label{1-9}  
\end{align}
\subsection{Lorentzian JT gravity and propagator}
Here we summarize how to describe the real time evolution of spatial slices in JT gravity. For more details see \cite{Late}. 

\begin{figure}[H]
\centering
\includegraphics[width=100mm]{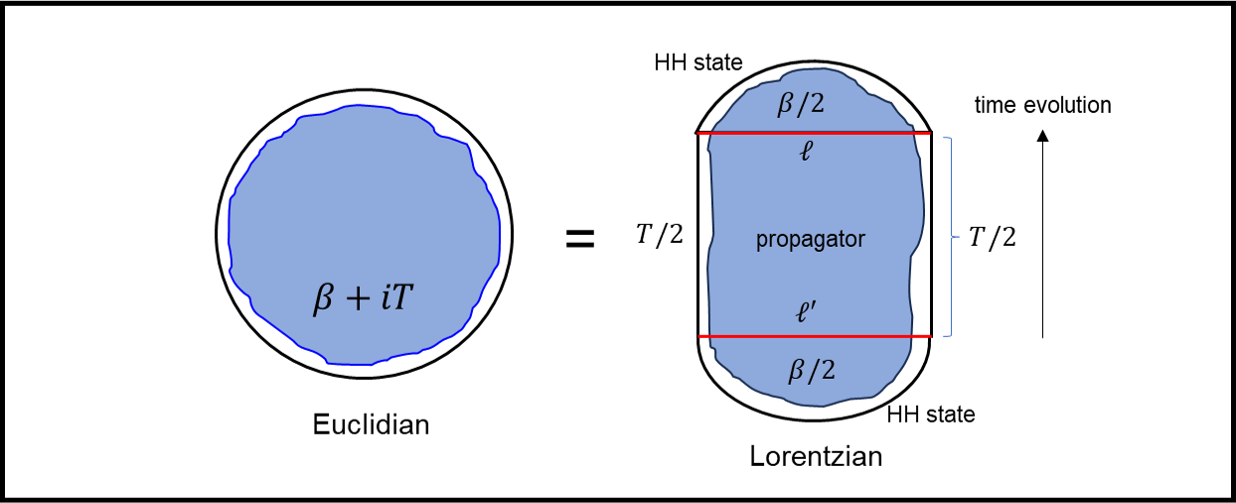}
\caption{\small (left): Disk amplitude with regularized boundary length $\beta + i T$. (right): The same amplitude made from two half-disks with Euclidean signature and  a rectangle with Lorentzian signature glued along the geodesics of lengths $\ell$ and $\ell'$}
\label{fig0-1(Lorentz)}
\end{figure}

The usual procedure involves first performing the path integral (\ref{1-5}) of the Euclidean theory, and then performing an analytic continuation to $\beta \rightarrow \beta \pm i T$. For example, the analytic continuation $Z(\beta) \to Z(\beta + iT)$ is equivalent to the geometry in the right of Fig.\ref{fig0-1(Lorentz)}, and it can be expressed as follows:\\
\begin{align}
Z(\beta+iT) &= \ \langle HH_\beta| e^{-i \frac{T}{2}(H_{\text{L}}+H_{\text{R}})} | HH_\beta\rangle \nonumber \\
&= \ \int^{\infty}_{-\infty} e^{\ell} d\ell \ \int^{\infty}_{-\infty} e^{\ell'} d\ell' \langle HH_\beta| \ell\rangle \langle \ell| e^{-i \frac{T}{2}(H_{\text{L}}+H_{\text{R}})} |\ell'\rangle \langle \ell'| HH_\beta\rangle, \label{1-10}
\end{align}
\noindent where $H_{L,R}$ are the Hamiltonians on the left and right boundaries and $|HH_{\beta}\rangle$ is the Hartle-Hawking (HH) state\footnote{This HH state describes the wormhole connecting two asymptotically AdS boundaries, allowing observers to view a black hole in equilibrium at inverse temperature $\beta$ from both sides of the wormhole \cite{HH-1,HH-2}.}. Its wavefunction in the length basis $|\ell \rangle$ can be expressed using the modified Bessel function $K_{\nu}(x)$ as follows \cite{Yang}:\\
\begin{align}
& \psi_{D,\beta/2}(\ell) \equiv \langle \ell | HH_\beta \rangle = \int_0^\infty dE \rho_0(E) e^{-\frac{\beta}{2} E} \ 4 e^{-\ell/2} K_{2i\sqrt{2E}} ( 4 e^{-\ell/2}), \label{1-11} \\ \nonumber
\end{align}
where $\rho_0(E) = \frac{e^{S_0}}{2\pi^{2}}\sinh(2\pi\sqrt{2E})=e^{S_0}\hat{\rho}_0(E)$ is the leading term of the state density in JT gravity. This HH wavefunction corresponds to the integral over all Euclidean surfaces of disk topology with the AdS boundary of regularized length $\beta/2$ and  a geodesic boundary\footnote{A geodesic boundary is a spatial boundary with zero extrinsic curvature, and is characterized by its regularized length. After introducing the holographic cutoff parameter $\epsilon$, the regularized length $\ell$ of the geodesic boundary is related to the bare length $\ell_b$ \cite{Harl2018} as:\begin{align} \ell \equiv \ell_b - 2 \log\bigg( \frac{2}{\epsilon}\bigg). \nonumber
\end{align} Thus $\ell$ is allowed to take negative values.} of length $\ell$. The modified Bessel function in the HH wavefunction is related to the bulk energy eigenstate $|E\rangle$  \cite{Late} as: \\
\begin{align}
& \psi_E(\ell)\equiv \langle \ell |E \rangle \equiv 4 e^{-\ell/2} K_{2i\sqrt{2E}} ( 4 e^{-\ell/2}) .\label{1-12} \\ \nonumber 
\end{align}
Thus, (\ref{1-11}) can also be rewritten as follows:\\
\begin{align}
& \psi_{D,\beta/2}(\ell)= \int_0^\infty  dE \rho_0(E) e^{-\frac{\beta}{2} E} \psi_E(\ell). \label{1-13} \\ \nonumber
\end{align}
$\langle \ell |E \rangle$ satisfies the following orthogonality and completeness relations:\\
\begin{align}
 e^{-S_0} \ \int^{\infty}_{-\infty} e^{\ell} d\ell \ \langle E_1 | \ell \rangle \langle \ell |E_2 \rangle & = \frac{\delta(E_1-E_2)}{\rho_0(E_1)}, \label{1-14} \\
 \int^{\infty}_{0} dE \ \rho_0(E) \langle \ell | E \rangle \langle E | \ell' \rangle & = \delta(\ell-\ell'). \label{1-15} \\ \nonumber 
\end{align}
The inner product of the HH wavefunctions corresponds to the disk partition function, as can be derived from (\ref{1-14}).

On the other hand, the analytic continuation $\beta \rightarrow \beta + i T$ of the HH wavefunction can be expressed using the propagator $P_{\chi=1}(T/2,\ell,\ell')$ as follows:\\
\begin{align}
\psi_{D,\beta/2+i T}(\ell) &= e^{-S_0}\int^{\infty}_{-\infty} e^{\ell'} d\ell' \ P_{\chi=1}(T/2, \ell , \ell') \psi_{D,\beta/2}(\ell'), \label{1-16} \\ \nonumber 
\end{align}
where $P_{\chi=1}(T/2, \ell, \ell')$ is the leading term of the propagator connecting the spatial slices with regularized lengths $\ell$ and $\ell'$, and corresponds to the rectangle in the right diagram of Fig.\ref{fig0-1(Lorentz)}. And, using (\ref{1-12}), $P_{\chi=1}(T/2, \ell, \ell')$ is explicitly given by:\\
\begin{align}
P_{\chi=1}(T/2,\ell,\ell') = \int^{\infty}_{0} dE \rho_0(E) e^{-i \frac{T}{2}E } \psi_E(\ell) \psi_E(\ell'). \label{1-17} \\ \nonumber
\end{align}
The subleading terms correspond to geometries with $g$ handles. The handles correspond to the emission and absorption of baby universes.

\begin{figure}[H]
\centering
\includegraphics[width=90mm]{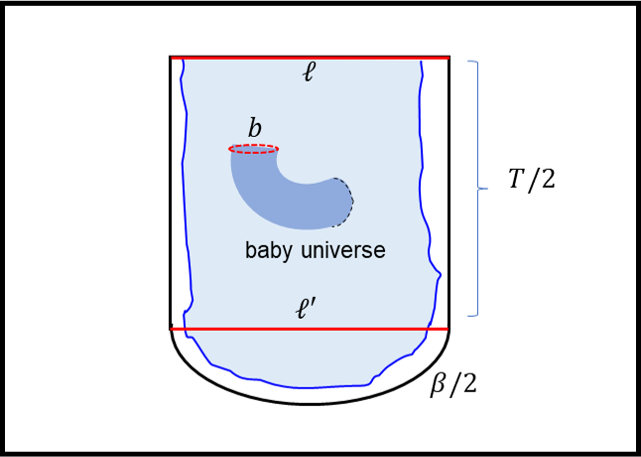} \caption{\small Contribution of the analytically continued trumpet wavefunction $\psi_{\text{Tr},\beta/2+i T}(\ell,b)$ (Lorentzian form)} 
\label{fig0-2(trum)}
\end{figure}

Similarly, the Lorentzian continuation of the trumpet partition function $Z_{\text{Tr}}(\beta,b)$, where $\beta$, $b$ are the lengths of the AdS and geodesic boundaries, can be expressed as follows \cite{Late}:\\ 
\begin{align}
Z_{\text{Tr}}(\beta+i T,b) &= e^{-S_0}\int^{\infty}_{-\infty} e^\ell \ d\ell \psi^*_{D,\beta/2}(\ell) \psi_{\text{Tr},\beta/2+i T}(\ell,b). \label{1-18} 
\end{align}
\\
Here the trumpet wavefunction $\psi_{\text{Tr},\beta/2}(\ell,b)$ is defined as \cite{Late}:\\ 
\begin{align}
\psi_{\text{Tr},\beta/2}(\ell,b) \equiv \langle \ell,b|HH_{\beta} \rangle &= \int_0^\infty dE \frac{\cos(b \sqrt{2 E})}{\pi \sqrt{2 E}} e^{-\frac{\beta}{2}E} \psi_E(\ell). \label{1-19} \\ \nonumber
\end{align}
For $T=0$, the trumpet wavefunction $\psi_{\text{Tr},\beta/2}(\ell,b)$ is given by the path integral over Euclidean geometries of cylindrical topology with a geodesic boundary of length $b$, an AdS$_2$ boundary of length $\beta/2$, and a geodesic boundary of length $\ell$. After analytic continuation $\beta/2 \rightarrow \beta/2 + iT$, the trumpet amplitude $Z_{\text{Tr}}(\beta+iT,b)$ can be interpreted as the amplitude for the HH state to transition into the state $|HH_\beta\rangle \otimes |b\rangle = |HH_\beta, b\rangle$ after a time $T/2$ has passed (see Fig.\ref{fig0-2(trum)}).

$Z_{\text{Tr}}(\beta+iT,b)$ can also be expressed as:\\
\begin{align}
Z_{\text{Tr}}(\beta+i T,b) &= e^{-S_0} \int^{\infty}_{-\infty} \prod_{i=1}^{2} e^{\ell_i} d\ell_i \langle HH_\beta |\ell_1 \rangle \langle \ell_1 ,b| e^{-i\frac{T}{2} (H_{\text{L}}+H_{\text{R}})} |\ell_2\rangle \langle \ell_2| HH_\beta \rangle \label{1-20} \\
& = e^{-S_0} \int^{\infty}_{-\infty} \prod_{i=1}^{2} e^{\ell_i} d\ell_i \psi^{*}_{D,\beta/2}(\ell_1) P_{\text{Tr}}(T/2,b,\ell_1,\ell_2) \psi_{D,\beta/2} (\ell_2). \label{1-21} 
\end{align}
\noindent The propagator $P_{\text{Tr}}(T/2, b, \ell, \ell') = \langle \ell, b| e^{-i\frac{T}{2} (H_{\text{L}}+H_{\text{R}})} |\ell'\rangle$ can be extracted from the trumpet wavefunction using the orthogonality relation (\ref{1-14}): \\
\begin{align}
P_{\text{Tr}}(T/2, b, \ell, \ell') = \int_0^\infty dE \frac{\cos(b \sqrt{2E})}{\pi\sqrt{2E}} e^{-i T E} \psi_E(\ell) \psi_E(\ell'). \label{1-22} \\ \nonumber 
\end{align}
This propagator can be decomposed into $P_{\chi=1}(T/2, \ell, \ell')$ and the amplitude $\langle \ell, b|\ell' \rangle$ as: \\
\begin{align}
P_{\text{Tr}}(T/2, b, \ell, \ell') &= e^{-S_0}\int^{\infty}_{-\infty} e^{\ell''} d\ell'' \;  \langle \ell, b| \ell''\rangle P_{\chi=1}(T/2, \ell'', \ell'),  \label{1-23} \\ \nonumber
\end{align}
where the amplitude $\langle \ell, b|\ell' \rangle$ is given by:\\
\begin{align}
\langle \ell,b|\ell' \rangle = \int^{\infty}_{0} dE\ \frac{\cos(b \sqrt{2E})}{\pi\sqrt{2E}} \psi_E(\ell) \psi_E(\ell').  \label{1-24} 
\end{align}

\section{Two-point correlation function} \label{Corre}
In this section, we explain how to calculate two-point correlation functions using the method described in \cite{Late,Volume}, which involves summing over all geodesics connecting the boundary operators. Then, to analyze the baby universe, we define an observable baby universe and distinguish different contributions to the two-point correlation functions between those which contain baby universes and those which do not. 
\subsection{The disk contribution}
\begin{figure}[H]
\centering
\includegraphics[width=130mm]{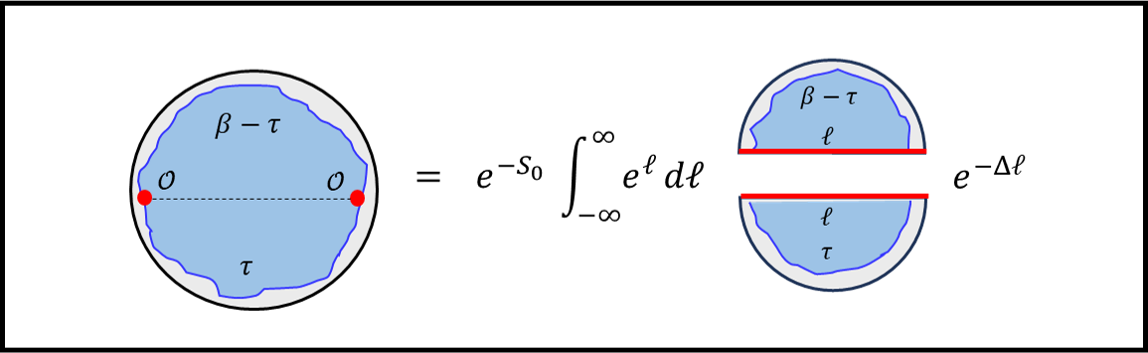}
\caption{\small Two-point correlation function on the disk in Euclidean geometry (\ref{2-3})}
\label{fig1-1(corr-1)}
\end{figure}

First, we consider the correlation function of two boundary operators on a disk with  Euclidean signature. Assuming that the operators do not have a direct interaction with the dilaton field, the correlation function is obtained by simply incorporating the correlation function of the matter theory into the path integral of JT gravity \cite{Yang}: \\
\begin{align} 
\langle \mathcal{O}(x_1) \mathcal{O}(x_2) \rangle_{\text{JT}} = \int \mathcal{D}F e^{-I_{\text{Sch}}(F(x))} \ \langle \mathcal{O}(x_1)\mathcal{O}(x_2) \rangle_{\text{CFT}}. \label{2-1} 
\end{align}
\\
When the boundary operators have conformal dimension $\Delta$, their correlation function is given by \cite{Yang}:\\
\begin{align}
\langle\mathcal{O}(x_1)\mathcal{O}(x_2)\rangle_{\text{CFT}}= e^{-\Delta \ell(x_1,x_2)}, \label{2-2} 
\end{align}
\\
where $\ell(x_1,x_2)$ represents the regularized geodesic distance between the boundary points $x_1$ and $x_2$. (\ref{2-1}) can be computed by first cutting the surface along the geodesic connecting the boundary points and then re-gluing (integrating) it, with an extra factor $e^{-\Delta \ell}$ as explained in \cite{Yang} (see Fig.\ref{fig1-1(corr-1)}): \\
\begin{align}
\langle\mathcal{O}(\tau)\mathcal{O}(0)\rangle_{\chi=1} &= e^{-S_0} \int^{\infty}_{-\infty} e^\ell \ d\ell \ \psi_{D,\beta-\tau} (\ell) \psi_{D, \tau}(\ell) e^{-\Delta \ell},  \label{2-3}  \\  
&= e^{-S_0} \int^{\infty}_{0} dE_1dE_2 \ \rho_0(E_1)\rho_0(E_2)\ e^{-(\beta-\tau)E_1-\tau E_2} |\mathcal{O}_{E_1,E_2}|^{2}, \label{2-4} \\ \nonumber
\end{align}
where we set $x_1 = \tau$ and $x_2 = 0$. Also, $\mathcal{O}_{E_1,E_2}$ is the matrix element of the operator in the energy basis, and is defined by the following $\ell$-integral:\\
\\
\begin{align}
|\mathcal{O}_{E,E'}|^2 \equiv \int_{-\infty}^\infty d\ell e^{\ell} \psi_E(\ell)\psi_{E'}(\ell) e^{-\Delta \ell} &= \frac{|\Gamma(\Delta+ i (\sqrt{2E}+\sqrt{2E'})) \Gamma(\Delta+ i (\sqrt{2E}-\sqrt{2E'}))|^2}{2^{2\Delta+1}\Gamma(2\Delta) }, \hspace{5pt} \Delta>0.  \label{2-5}
\end{align}
\noindent Depending on whether we are interested in the two-sided or one-sided correlation function, we analytically continue $\tau$ in (\ref{2-4}) to $\beta/2 + iT$ or $iT$. To be more explicit, the two-sided correlation function reads\\
\begin{align}
\langle\mathcal{O}(\beta/2+iT)\mathcal{O}(0)\rangle_{\chi=1}&= e^{-S_0} \int^{\infty}_{-\infty} e^{\ell} d\ell \ |\psi_{D,\beta/2+ i T}(\ell)|^2 e^{-\Delta \ell} \nonumber \\
&= e^{-S_0} \int^{\infty}_{0} dE_1dE_2 \ \rho_0(E_1)\rho_0(E_2)\ e^{-\beta(E_1+E_2)-iT(E_1-E_2)} |\mathcal{O}_{E_1,E_2}|^{2}. \label{2-6} 
\end{align}
\\
This Lorentzian correlation function can be interpreted as the expectation value of $e^{-\Delta \ell}$ in the HH state which has time evolved for $T/2$.
\subsection{The first subleading contribution}
\begin{figure}[H]
\centering
\includegraphics[width=160mm]{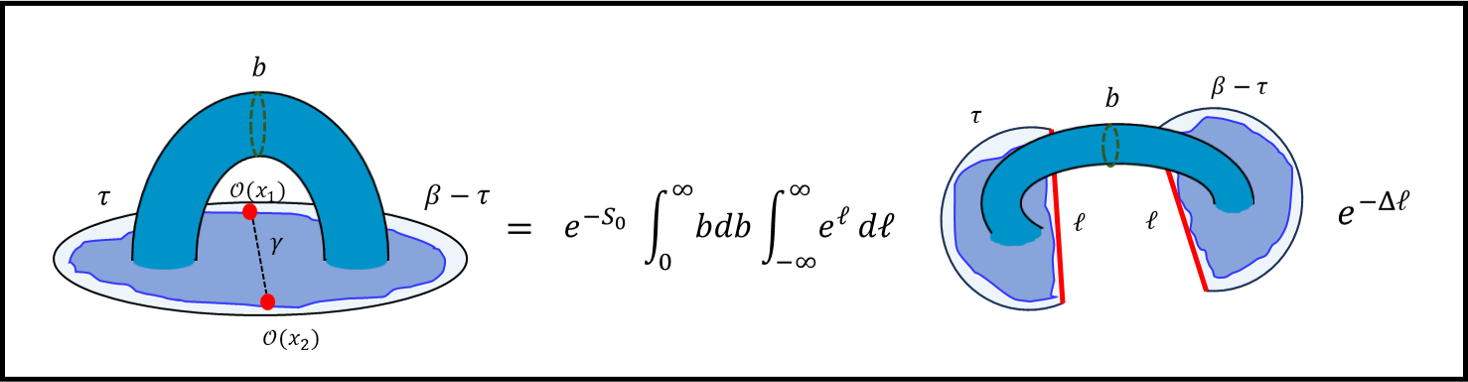} 
\caption{\small $g=1$ contribution to the two-point correlation function}
\label{fig1-3(ramp-1)}
\end{figure}

\begin{figure}[H]
\centering
\includegraphics[width=140mm]{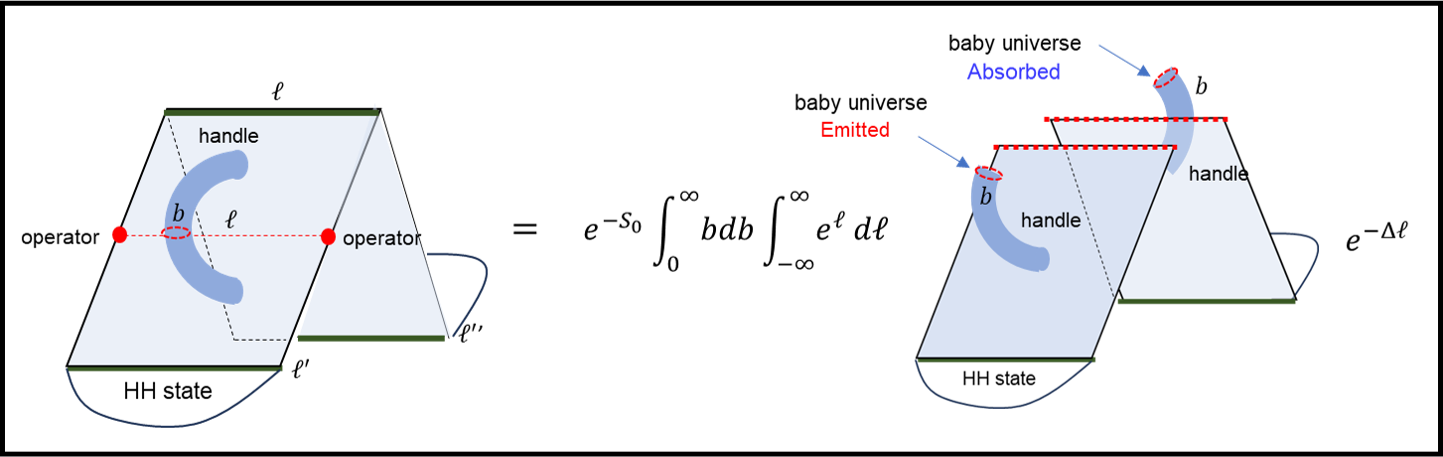}
\caption{\small The two-sided two-point correlation function from a disk with one handle (Lorentzian signature) includes a ramp from the contribution of geodesic passing through the handle. This arises from the parent universe emitting and absorbing a baby universe at late times $T$.}
\label{fig1-4(ramp-2)}
\end{figure}

Let's next consider the correlation function on a disk with one handle. In this case, a new complication arises because there are infinitely many geodesics connecting $x_1$ and $x_2$ that are homotopically inequivalent. According to \cite{Late,Volume} the CFT correlators in such cases are given by summing over all those geodesics:\\
\begin{align}
\langle \mathcal{O}(x_1)\mathcal{O}(x_2) \rangle_{\text{CFT}} = \sum_{\gamma \in \mathcal{G}_{x_1,x_2}} e^{-\Delta \ell_\gamma}, \label{2-7} 
\end{align}
\\
where $\mathcal{G}_{x_{1},x_{2}}$ denotes the set of all geodesics connecting $x_{1}$ and $x_{2}$. Of particular importance are the terms for which $\gamma$ is chosen to ``pass through the handle''. For such case, cutting along $\gamma$ results in a connected hyperbolic surface with one less genus (double trumpet) as in Fig.\ref{fig1-3(ramp-1)}. By summing over such $\gamma$ one obtains \cite{Late}:\\
\begin{align}
\langle\mathcal{O}(\tau)\mathcal{O}(0)\rangle_{\chi=-1}& \supset \ e^{-S_{0}} \int^{\infty}_{-\infty} \ e^{\ell}e^{-\Delta \ell}\ d\ell \ \int^{\infty}_{0} db \int^{b}_{0} d\tau  \ \psi_{\text{Tr},\beta-\tau}(\ell,b) \ \psi_{\text{Tr},\tau}(\ell,b)  \nonumber \\
&=  \ e^{-S_{0}} \int^{\infty}_{0} bdb \int^{\infty}_{0}\ dE_1dE_2 \ \rho_{\text{Tr}}(E_1,b) \ \rho_{\text{Tr}}(E_2,b)  \ e^{-(\beta-\tau)E_1}\ e^{-\tau E_2}|\mathcal{O}_{E_1,E_2}|^{2},  \label{2-8}  \\ \nonumber
\end{align}
where $\rho_{\text{Tr}}(E,b) = \frac{\cos(b \sqrt{2E})}{\pi\sqrt{2E}}$. By analytic continuation one obtains a geometry with Lorentzian signature as in Fig.\ref{fig1-4(ramp-2)}.

The contribution (\ref{2-8}) to the correlation function gives rise to the ``ramp'' behavior, i.e. the linear growth in $T$ at late times $T \sim e^{S_0/2}$. As reviewed in Appendix \ref{AppendixSFF}, this kind of behavior was first found in the analysis of SFF.
The derivation of (\ref{2-8}) involves some detailed comparison of the moduli space of surfaces before and after cutting along the chosen geodesic $\gamma$. This has been discussed in great detail in \cite{Volume} for each term in the genus expansion. Here we would like to review their argument and expand on them a bit.

The genus $g$ contribution to the two-point correlation function can be expressed as:\\
\begin{align}
\langle \mathcal{O}(x_1)\mathcal{O}(x_2) \rangle_{\chi=1-2g}=&  e^{S_0(1-2g)} \int^{\infty}_{-\infty}d\ell \int^{\infty}_{0}db_{\text{Tr}} \int^{b_{\text{Tr}}}_{0}d\tau_{\text{Tr}} \int_{\text{Mod}(\mathcal{M}_{g,1})} \omega \int D{(\mathcal{W})} e^{-I_\partial(b_\text{Tr}, \mathcal{W})}  \nonumber \\
& \hspace{25pt} \times \sum_{\gamma \in \mathcal{G}_{x1, x2}} e^{-\Delta \ell_{\gamma}} \delta(\ell-\ell_{\gamma}(\text{moduli})), \label{2-9}
\end{align}
\\
where $\mathcal{D}(\mathcal{W})$ and $\omega$ stand for the integration measures for the boundary wiggles and the bulk moduli, and $\ell_{\gamma}(\text{moduli})$ is the length of a geodesic that connects the boundary operators, which should be some function of the bulk moduli. $b_{\text{Tr}}$ and $\tau_{\text{Tr}}$ are the length/twist parameters of the circle between a trumpet and a surface with genus $g$ and $n=1$.

The moduli space $\text{Mod}(\mathcal{M}_{g,n})$ of surfaces with genus $g$ and $n$ geodesic boundaries of length $b_1,\cdots,b_n$ is defined as follows. Let $\mathcal{G}(\mathcal{M}_{g,n})$ be the space consisting of all values of the metric tensor on a hyperbolic surface $\mathcal{M}_{g,n}$. The moduli space $\text{Mod}(\mathcal{M}_{g,n})$ is then defined as the following quotient space \cite{corre-1, JTRandom, JTRandom-2}:\\
\begin{align}
\text{Mod}(\mathcal{M}_{g,n}) := \mathcal{G}(\mathcal{M}_{g,n})/\text{Diff}(\mathcal{M}_{g,n}), \label{2-10}
\end{align}
\\
where $\text{Diff}(\mathcal{M}_{g,n})$ denotes the group of diffeomorphisms on the surface $\mathcal{M}_{g,n}$. Note that in JT gravity, the curvature of $\mathcal{M}_{g,n}$ is fixed by the equations of motion of the dilaton, which is equivalent to dividing by the volume associated with the Weyl transformation, as we do in string theory. So the $\text{Mod}(\mathcal{M}_{g,n})$ becomes finite-dimensional by dividing only by the volume associated with the diffeomorphism.

There is a similar notion called Teichmüller space $\mathcal{T}_{g,n}$ which is defined as follows \cite{Witten:1990hr, Teich}:\\
\begin{align}
\mathcal{T}_{g,n} := \mathcal{G}(\mathcal{M}_{g,n})/\text{Diff}_{0}(\mathcal{M}_{g,n}), \label{2-11}
\end{align}
\\
where $\text{Diff}_0(\mathcal{M}_{g,n})$ represents the group of diffeomorphisms isotopic to the identity. $\text{Mod}(\mathcal{M}_{g,n})$ can be represented as the quotient of $\mathcal{T}_{g,n}$ by the mapping class group $\text{MCG}(\mathcal{M}_{g,n})$:\\
\begin{align}
\text{Mod}(\mathcal{M}_{g,n}) :=\ & \mathcal{T}_{g,n}/\text{MCG}(\mathcal{M}_{g,n}), \label{2-12} \\
\text{MCG}(\mathcal{M}_{g,n}) :=\ & \text{Diff}(\mathcal{M}_{g,n})/\text{Diff}_0(\mathcal{M}_{g,n}). \label{2-13} \\ \nonumber 
\end{align}

The standard measure $\omega$ for the bulk moduli is constructed as follows. As is well-known, metrics on a hyperbolic surface $\mathcal{M}_{g,n}$ of genus $g$ with $n$ geodesic boundaries can be parametrized by $2(3g+n-3)$ length and twist parameters $(\vec{b}, \vec{\tau}) \in \mathbb{R}^{3g+n-3}_{+} \times \mathbb{R}^{3g+n-3}$, and these parameters can be used as a coordinate on the moduli space (called Fenchel-Nielsen coordinates). Using them, the Weil-Petersson symplectic form $\Omega$ can be expressed as follows \cite{mirzakhani2007simple, Mirzakhani:2006eta, Witten:1990hr, Kontsevich:1992ti,JTRandom}:\\
\begin{align}
\Omega = \sum_{i=1}^{3g-3+n}db_{i} \wedge d\tau_{i}, \label{2-14}
\end{align}
\\
and the measure is given by $\omega=\frac{\Omega^{3g+n-3}}{(3g+n-3)!}$.\\

Now let's express (\ref{2-9}) as an integral over the moduli of the surface cut along the geodesic $\gamma$. For convenience, we denote by $\mathcal{\hat{M}}_{g,1}$ the original surface containing the trumpet part, of which the moduli space has dimension:\\
\begin{align}
\text{dim}\big(\text{Mod}(\mathcal{\hat{M}}_{g,1})\big) = \text{dim}\big( \text{Mod}(\mathcal{M}_{g,1}) \big)+2 = 6g-2, \label{2-15}\\ \nonumber  
\end{align}
and the surface cut along $\gamma$ will be denoted as $\mathcal{\hat{M}}_{g,1}\setminus \gamma$.

Depending on the choice of $\gamma$, the cut geometry is either a connected surface of genus $g-1$ with two asymptotic boundaries, or two disconnected surfaces of genus $h$ and $g-h$ each having one asymptotic boundary. So, $\text{Mod}(\mathcal{\hat{M}}_{g,1} \setminus \gamma)$ is parametrized by the moduli of $\mathcal{M}_{g-1,2}$ or $\mathcal{M}_{h,1} \cup \mathcal{M}_{g-h,1}$ along with the length/twist parameters $b_i, \tau_i (i=1,2)$ for attaching two trumpets. Its dimension is\\ 
\begin{align}
\text{dim}\big(\text{Mod}(\mathcal{\hat{M}}_{g,1}\setminus \gamma)\big) &= \{ \text{dim}\big( \text{Mod}(\mathcal{M}_{g-1,2}) \big)+4 \quad \text{or} \quad \text{dim}\big( \text{Mod}(\mathcal{M}_{h,1}) \big) + \text{dim}\big( \text{Mod}(\mathcal{M}_{g-h,1}) \big)+4 \} \nonumber \\
&= 6g-4. \label{2-16}
\end{align}
\\
Comparing this with (\ref{2-15}) we see that the dimension is reduced by two although we have introduced only one condition $\ell=\ell_{\gamma}(\text{moduli})$.

We recall here that the spaces $\text{Mod}(\mathcal{\hat{M}}_{g,1})$ and $\text{Mod}(\mathcal{\hat{M}}_{g,1}\setminus \gamma)$ both have a symplectic structure. Therefore, it must be that $\text{Mod}(\mathcal{\hat{M}}_{g,1}\setminus \gamma)$ is the symplectic reduction of $\text{Mod}(\mathcal{\hat{M}}_{g,1})$ by the Hamiltonian action generated by the moment map $\ell_{\gamma}(\text{moduli})$. Namely, fixing the length of the geodesic $\gamma$ via the moment map condition reduces the dimension by one, and a further quotient by the Hamiltonian flow generated by $\ell_{\gamma}$ reduces the dimension by one, which results in a total reduction of two dimensions. Note that the points along an orbit of the flow correspond to surfaces with different shapes before cutting along $\gamma$, but after the cut they all have the same shape since they all correspond to the same point on $\text{Mod}(\mathcal{\hat{M}}_{g,1}\setminus \gamma)$.

Thus the integral over $\text{Mod}(\mathcal{\hat{M}}_{g,1})$ with $\ell=\ell_{\gamma}(\text{moduli})$ fixed can be decomposed into those over $\text{Mod}(\mathcal{\hat{M}}_{g,1}\setminus \gamma)$ and along the flow. 
In order for the simple dimensional reduction of \cite{Volume} to work, it must be that the integral along the flow (volume of the orbit) is independent of $\ell$ and the moduli of $\text{Mod}(\mathcal{\hat{M}}_{g,1}\setminus \gamma)$.

Having understood the relation between the local structure of $\text{Mod}(\mathcal{\hat{M}}_{g,1})$ and $\text{Mod}(\mathcal{\hat{M}}_{g,1}\setminus \gamma)$, we now turn to the role of the MCGs. As explained above, the geodesics $\gamma$ between boundary points of $\mathcal{\hat{M}}_{g,1}$ can be classified according to whether one obtains $\mathcal{\hat{M}}_{g-1,2}$ or $\mathcal{\hat{M}}_{h,1} \cup \mathcal{\hat{M}}_{g-h,1}$ for some $h$ by cutting $\mathcal{\hat{M}}_{g,1}$ along $\gamma$. The MCG of $\mathcal{\hat{M}}_{g,1}$ transforms the boundary-to-boundary geodesics within each cutting class among themselves, as well as transforming the closed geodesics among themselves. the subgroup of $\text{MCG}(\mathcal{\hat{M}}_{g,1})$ which acts on a given $\gamma$ trivially is equal to the MCG of the surface that has been cut along $\gamma$. Therefore, dividing by $\text{MCG}(\mathcal{\hat{M}}_{g,1})$ and summing over $\gamma$ in a given cutting class is equivalent to just dividing by $\text{MCG}(\mathcal{\hat{M}}_{g-1,2})$ or $\text{MCG}(\mathcal{\hat{M}}_{h,1}) \otimes \text{MCG}(\mathcal{\hat{M}}_{g-h,1})$.

Thus, by combining what has been explained in the above, we find \cite{Volume}\\
\begin{align}
& \int^{\infty}_{-\infty}d\ell \int^{\infty}_{0}db_{\text{Tr}} \int^{b_{\text{Tr}}}_{0}d\tau_{\text{Tr}}  \int_{\text{Mod}(\mathcal{M}_{g,1})}  \omega \sum_{\gamma \in \mathcal{G}_{x1, x2}} e^{-\Delta \ell_{\gamma}} \delta(\ell-\ell_{\gamma}(\text{moduli})) \nonumber \\
&\hspace{25pt} =\int^{\infty}_{-\infty}e^{\ell} d\ell \int \prod^{2}_{i=1}db_i d\tau_i \ \bigg[ e^{-\Delta \ell} \int_{\text{Mod}(\mathcal{M}_{g-1,2})}  \omega + \sum_{h \geq 0} e^{-\Delta \ell}\int_{\text{Mod}(\mathcal{M}_{h,1})}  \omega \int_{\text{Mod}(\mathcal{M}_{g-h,1})}  \omega \ \bigg]. \label{2-17} 
\end{align}
\\
\\
The integral over each moduli space $\text{Mod}(\mathcal{M}_{g,n})$ gives the Weil-Petersson volume $V_{g,n}(b_1,\cdots,b_n)$ \cite{mirzakhani2007simple,Mirzakhani:2006eta,Kontsevich:1992ti,Eynard2007}. Thus the two-point correlation function (\ref{2-9}) can be rewritten as follows:\\
\begin{align}
\langle \mathcal{O}(\tau)\mathcal{O}(0) \rangle_{\chi=1-2g} &\sim e^{S_0(1-2g)}\ \int^{\infty}_{-\infty}  e^\ell d\ell \int \prod^{2}_{i=1} \ db_i d\tau_i \ \psi_{\text{Tr},\tau}(\ell, b_1) \psi_{\text{Tr},\beta - \tau}(\ell, b_2) e^{-\Delta \ell} \nonumber \\ 
&\hspace{20pt} \times \bigg[ V_{g-1, 2}(b_1, b_2)  +  \sum_{h\geq 0}  V_{g-h, 1}(b_1)  V_{h,1}(b_2) \bigg]. \label{2-18} \\ \nonumber
\end{align}
In particular, the $g=1(\chi=-1)$ contribution is given by:\\
\begin{align}
\langle \mathcal{O}(\tau)\mathcal{O}(0) \rangle_{\chi=-1} & \sim e^{-S_0} \int^{\infty}_{-\infty}  e^\ell d\ell \int \prod^{2}_{i=1} \ db_i d\tau_i  \ \psi_{\text{Tr},\tau}(\ell, b_1) \psi_{\text{Tr},\beta - \tau}(\ell, b_2) e^{-\Delta \ell} \nonumber \\ 
& \times \bigg[ V_{0, 2}(b_1, b_2)  +  V_{0,1}(b_1)  V_{1,1}(b_2) + V_{1,1}(b_1)  V_{0,1}(b_2) \bigg] \nonumber \\
& = e^{-S_0} \int^{\infty}_{0} b db\ \int dE_1dE_2\ \rho_{\text{Tr}}(E_1,b)\rho_{\text{Tr}}(E_2,b)e^{-(\beta-\tau)E_1-\tau E_2}|\mathcal{O}_{E_1,E_2}|^{2} \label{2-19} \\
& +e^{-S_0} \int^{\infty}_{0}  bdb \ V_{1,1}(b) \ \int dE_1dE_2 \ \rho_{0}(E_1)\rho_{\text{Tr}}(E_2,b) e^{-(\beta-\tau)E_1-\tau E_2}|\mathcal{O}_{E_1,E_2}|^{2} \label{2-20} \\
& + e^{-S_0} \int^{\infty}_{0} bdb \ V_{1,1}(b) \ \int dE_1dE_2 \ \rho_{\text{Tr}}(E_1,b)\rho_{0}(E_2) e^{-(\beta-\tau)E_1-\tau E_2}|\mathcal{O}_{E_1,E_2}|^{2}, \label{2-21} \\ \nonumber
\end{align}
The contribution from $V_{0, 2}(b_1, b_2)$ in (\ref{2-19}) arises from the situation in which the geodesic $\gamma$ passes through the handle, which brings about the ramp as seen in (\ref{2-8}). The remaining terms include $V_{0, 1}(b)$ which cannot be obtained from Mirzakhani's recursion relation \cite{mirzakhani2007simple,Mirzakhani:2006eta} . We have set $V_{0, 1}(b)$ so that the disk wave function is reproduced after the $b$ integral \cite{Volume}:\\
\begin{align}
\psi_{D,x}(\ell) = \int_0^\infty dE \rho_{0}(E) \psi_{E}(\ell)e^{- x E} \equiv \int_0^\infty db b \ \psi_{\text{Tr},x}(\ell, b) V_{0, 1}(b). \label{2-22}
\end{align}
\subsection{\normalsize The presence or absence of baby universes}
\begin{figure}[H]
\centering
\includegraphics[width=140mm]{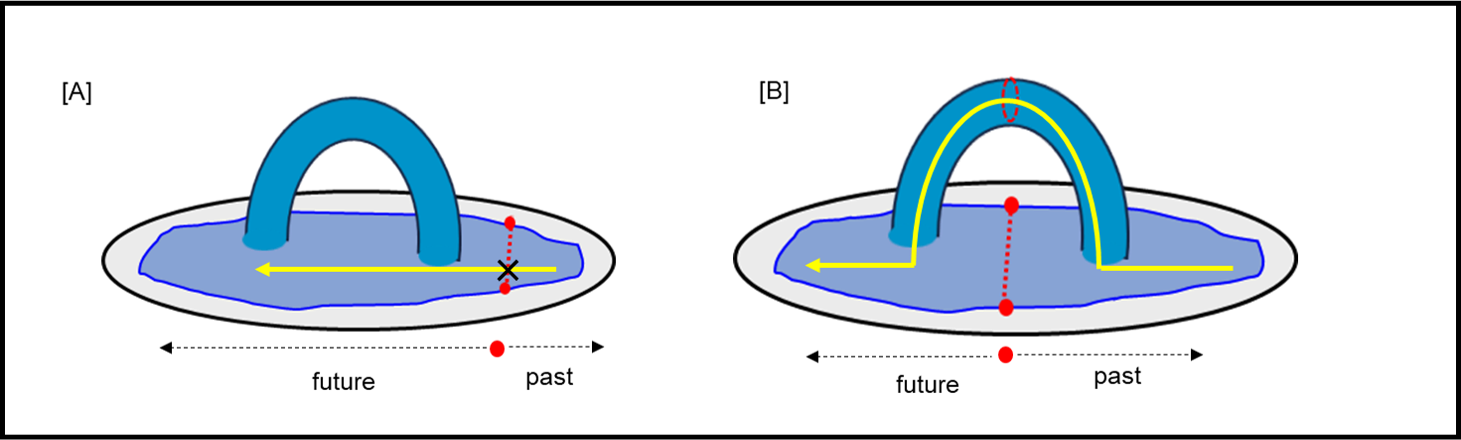}
\caption{\small [A] a particle worldline intersecting the equal-time slice. [B] the worldline can reach the future from the past without intersecting the equal-time slice, by passing through the handle.} 
\label{fig1-5(baby-universe)} 
\end{figure}

It is natural to regard the geodesic $\gamma$ connecting the two boundary operators as the spatial slice at time $T$. Then, whether there is an observable baby universe at time $T$ is determined by the homotopy type of $\gamma$. If $\gamma$ ``passes through a handle'', then cutting the surface along $\gamma$ results in a single connected surface with one less genus. Then an observable baby universe allows particles to bypass $\gamma$ when traversing from boundary points with time $<T$ to those with time $>T$ as shown in Fig.\ref{fig1-5(baby-universe)} [B].

In (\ref{2-18}), the first term in the square bracket corresponds to the contribution from this kind of situation. By summing over $g$ one has \\ 
\begin{align}
& \langle \mathcal{O}(\beta/2+iT) \mathcal{O}(0)\rangle_{\text{baby}} \sim e^{-S_0}\int^{\infty}_{0} dE_1 dE_2 \ e^{-\frac{\beta}{2}(E_1+E_2)+iT(E_1-E_2) } \rho_{\text{conn}}(E_1,E_2) |\mathcal{O}_{E_1,E_2}|^{2}, \label{2-23} \\ \nonumber
\end{align}
where we set $x_1 \to \beta/2+iT, x_2 \to 0$, and $\rho_{\text{conn}}(E_1,E_2)$ is defined by: \\
\begin{align}
\rho_{\text{conn}}(E_1,E_2) \equiv \sum^{\infty}_{g=1} e^{S_0(1-2g)} \int^{\infty}_{0} \prod_{i=1}^{2}\ b_i db_i \ \rho_{\text{Tr}}(E_1,b_1) \rho_{\text{Tr}}(E_2,b_2) V_{g-1,2}(b_1,b_2). \label{2-24} 
\end{align}
\noindent The term for $g=1$ corresponds to (\ref{2-8}) or (\ref{2-19}). On the other hand, if cutting the surface along $\gamma$ gives two disconnected surfaces, then there is no baby universe at time $T$. This situation corresponds to the second term in the square bracket of (\ref{2-18}), and by summing over $g$ one obtains \\
\begin{align} 
\langle \mathcal{O}(\beta/2+iT) \mathcal{O}(0)\rangle_{\text{no-baby}} \sim e^{-S_0}\int^{\infty}_{0} dE_1 dE_2 \ e^{-\frac{\beta}{2}(E_1+E_2)+iT(E_1-E_2) } \rho(E_1)\rho(E_2)|\mathcal{O}_{E_1,E_2}|^{2}, \label{2-25}\\ \nonumber
\end{align}
where $\rho(E)$ is defined by: \\
\begin{align}
\rho(E) \equiv e^{S_0}\hat{\rho}_{0}(E) + \sum_{g=1}^{\infty} e^{S_0(1-2g)} \int^{\infty}_{0} bdb \ \rho_{\text{Tr}}(E,b) V_{g,1}(b). \label{2-26} \\ \nonumber
\end{align}

In this section, we calculated the two-point correlation functions according to the method of \cite{Late,Volume}, but there are other natural approaches. Among them, we are interested in the following two.

The first is the approach of \cite{Firewall-1} which evaluated the handle-disk contribution to the JT partition function. In that work, one first selects a specific geodesic connecting two boundary points as the spatial slice, and then integrates over a certain region of the moduli space determined by the ``no-shortcut'' condition which ensures that the chosen geodesic is the shortest.  Thus, as compared to \cite{Late,Volume} it differs in the region of integration over moduli parameters $b, \tau$. 

The other approach is to use the formula for the universal spectral density correlation of random matrices in the $\tau$-scaling limit. Based on this approach, the behavior of correlation functions, especially at late times, can be analyzed accurately without non-perturbative information \cite{SSS,plateau1}.

It is interesting to see how these three approaches differ. With regard to the first two perturbative methods, the difference is whether to take the shortest geodesic connecting the two boundary operators or the sum of an infinite set of homotopically inequivalent geodesics. So the difference in the values of correlators is roughly an integral of $\sim e^{-\Delta \ell_{\text{2nd}}}$ over the moduli space, where $\ell_{\text{2nd}}$ is the second shortest geodesic. Since $\ell_{\text{2nd}}$ should be much longer than $\ell_{\text{1st}}$ at $T\gg1$ at generic points on the moduli space, we expect that the difference is suppressed at late times.

Note also that SFF analysis suggests that at late times, the contributions to the correlation functions with one observable baby universe should consistently exhibit ramp behavior across all the three methods. In the next section, we calculate the correlation function explicitly and examine the above expectations.

\section{Calculation of the correlation function} \label{Corr-sin}
In this section, we compute the correlation function following the three approaches mentioned above, and compare the late times behavior. First, we review the method of \cite{Firewall-1} and compute the correlation function. For the contribution from geometries without an observable baby universe, the disk geometry ($g=0$) is the leading term in (\ref{2-25}). For the contribution from geometries with an observable baby universe, we focus on the cylinder geometry ($g=1$), which is the leading term in (\ref{2-23}). Then we also compute the correlation function in the $\tau$-scaling limit \cite{OS} using the universality of random matrix theory and the spectral density correlation formula.  Finally, we compare the results obtained by the three methods, particularly focusing on their late times behavior.

\subsection{\normalsize Disk ($g=0$) contribution}
\begin{figure}[H]
\centering
\includegraphics[width=70mm]{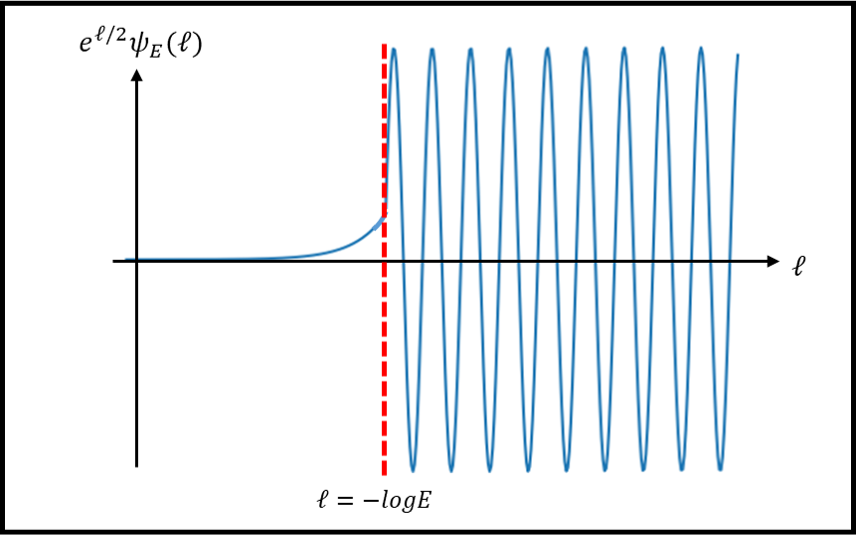}
\caption{\small Graph of (\ref{3-6})}
\label{fig3-1(cos)}
\end{figure}

First, we extract the disk contribution to the correlation function (\ref{2-23}). After the analytic continuation ($x_1 \to \beta/2+iT, x_2 \to 0$), it is given by:\\
\begin{align}
\langle \mathcal{O}(x_1) \mathcal{O}(x_2)\rangle_{\text{no-baby}}^{\chi=1}  &= e^{-S_0} \int^{\infty}_{-\infty}e^{\ell}d\ell e^{-\Delta \ell} \int^{\infty}_{0} dE_1dE_2 \ e^{-\frac{\beta}{2}(E_1+E_2)-iT(E_1-E_2) } \rho_0(E_1)\rho_0(E_2) \psi_{E_1}(\ell)\psi_{E_2}(\ell) \label{3-1} \\
& = e^{-S_0} \int^{\infty}_{-\infty}e^{\ell}d\ell e^{-\Delta \ell} \ \psi_{D,\beta/2+iT}(\ell)\psi_{D,\beta/2-iT}(\ell). \label{3-2} 
\end{align}
\\
We use the inverse Laplace transform to fix the energy $E$:\\
\begin{align}
\langle \mathcal{O}(x_1) \mathcal{O}(x_2)\rangle_{\text{no-baby}}^{\chi=1} = e^{-S_0}\int^{\infty}_{-\infty}e^{\ell}d\ell e^{-\Delta \ell}  \int_{\mathcal{C}}  \frac{d\beta}{2\pi i} e^{\beta E} \psi_{D,\beta/2+iT}(\ell)\psi_{D,\beta/2-iT}(\ell). \label{3-3} \\ \nonumber  
\end{align}
Since the $\beta$ integral gives $E_1 + E_2 = 2E$, we set the energies $E_1$ and $E_2$ as follows: \\
\begin{align}
E_1=E+\frac{\omega}{2}, \hspace{20pt} E_2=E-\frac{\omega}{2}. \label{3-4}
\end{align}
\\
Then, (\ref{3-3}) can be rewritten as follows: \\
\begin{align}
\langle \mathcal{O}(x_1) \mathcal{O}(x_2)\rangle_{\text{no-baby}}^{\chi=1} = e^{-S_0}\int^{\infty}_{-\infty}e^{\ell}d\ell e^{-\Delta \ell}  \int^{\infty}_{-\infty} d\omega e^{-iT\omega } \rho_0(E_1)\rho_0(E_2)\psi_{E_1}(\ell)\psi_{E_2}(\ell). \label{3-5} 
\end{align}
\\
To evaluate this integral, we assume $E, T \gg 1$ and apply the semiclassical approximation used in \cite{Firewall-1}. Under this assumption, the modified Bessel function $K_{2i\sqrt{2E}}(4e^{-\ell/2})$ contained in $\psi_{E}(\ell)$ behaves as in Fig.\ref{fig3-1(cos)}. It is exponentially suppressed in the region $\ell < -\log E$, while it oscillates rapidly in the region $\ell > -\log E$. In this oscillatory region $K_{2i\sqrt{2E}}(4e^{-\ell/2})$ can be approximated as follows \cite{bessal}:\\ 
\begin{align}
&e^{\ell/2}\psi_{E}(\ell) = 4K_{2i\sqrt{2E}}(4e^{-\ell/2}) \approx \frac{4\pi^{1/2}}{(2E)^{1/4}} e^{-\pi\sqrt{2E}} \cos \left[\ \sqrt{2E} (\ell+\log(2E)-2) -\frac{\pi}{4} \ \right]. \label{3-6}
\end{align}
Considering (\ref{3-6}) and $E\gg1$ in $\ell >-\log E$, (\ref{3-5}) can be rewritten as follows: \\
\begin{align}
&\langle \mathcal{O}(x_1) \mathcal{O}(x_2)\rangle_{\text{no-baby}}^{\chi=1} \nonumber \\
&\approx \ \frac{e^{S_0+2\pi\sqrt{2E}}}{2\pi^2(2E)^{1/2}}\int^{\infty}_{-\log E} d\ell e^{-\Delta \ell}  \int^{\infty}_{-\infty} \frac{d\omega}{2\pi} e^{-iT\omega } \left( e^{2i\sqrt{2E}L-\frac{i\pi}{2}}+e^{\frac{i\omega}{\sqrt{2E}}L} + e^{-\frac{i\omega}{\sqrt{2E}}L} + e^{-2i\sqrt{2E}L+\frac{i\pi}{2}} \right), \label{3-7} \\ \nonumber
\end{align}
where we set $L=\ell+\log(2E)-2$. Also, $e^{2i\sqrt{2E}L-\frac{i\pi}{2}}$ and $e^{-2i\sqrt{2E}L+\frac{i\pi}{2}}$ oscillate rapidly under the assumption $E \gg 1$, so their oscillation phases average out to zero. Thus, two delta functions are generated from the $\omega$ integral as follows:\\
\begin{align}
\langle \mathcal{O}(x_1) \mathcal{O}(x_2)\rangle_{\text{no-baby}}^{\chi=1} &\approx \frac{e^{S_0+2\pi\sqrt{2E}}}{2\pi^2(2E)^{1/2}}\int^{\infty}_{-\log E} d\ell e^{-\Delta \ell}  \int^{\infty}_{-\infty} \frac{d\omega}{2\pi} \ \bigl(\ e^{-i\left(T-\frac{iL}{\sqrt{2E}}\right)} \ + \ e^{-i\left(T+\frac{iL}{\sqrt{2E}}\right)}  \ \bigr) \label{3-8} \\
& = \frac{e^{S_0+2\pi\sqrt{2E}}}{2\pi^2(2E)^{1/2}}\int^{\infty}_{-\log E} d\ell e^{-\Delta \ell}  \ \left[ \ \delta \bigg(T-\frac{L}{\sqrt{2E}}\bigg) \ +\  \delta \bigg(T+\frac{L}{\sqrt{2E}} \bigg) \  \right]. \label{3-9} \\ \nonumber
\end{align}
Since $T \gg 1$, the delta function in the second term can be dropped, and as a result, the disk contribution becomes\\
\begin{align}
\langle \mathcal{O}(x_1) \mathcal{O}(x_2)\rangle_{\text{no-baby}}^{\chi=1} \approx \frac{e^{S_0+2\pi\sqrt{2E}}}{2\pi^2} \int^{\infty}_{-\log E} d\ell e^{-\Delta \ell} \delta(\ell-\ell_T) = \frac{e^{S_0+2\pi\sqrt{2E}}}{2\pi^2} e^{-\Delta \ell_T}, \label{3-10} \\ \nonumber
\end{align}
where we set $\ell_T = \sqrt{2E}T - \log(2E) + 2$ and used $\ell_T \gg -\log E$. As a result, the disk contribution to the correlation function concentrates around a specific geodesic length $\ell_T = \sqrt{2E}T - \log(2E) + 2 \sim \sqrt{2E}T$ and decays exponentially in time.

Finally, by applying the Laplace transform to (\ref{3-10}), we obtain the following function of $\beta$:\\
\begin{align}
\langle \mathcal{O}(x_1) \mathcal{O}(x_2)\rangle_{\text{no-baby}}^{\chi=1(\beta)} &\approx \ \frac{e^{S_0}}{2\pi^{2}} \int^{\infty}_{0} \ dE e^{-\beta E} e^{-(\Delta T-2\pi)\sqrt{2E}} \nonumber \\
& = e^{S_0+\frac{(\Delta T-2\pi)^{2}}{2\beta}}\ \left[ \ \frac{1}{2\pi^{2} \beta} -\frac{(\Delta T-2\pi)}{(2\pi \beta)^{3/2}}\ \left( 1- \text{Erf}\left( \frac{(\Delta T-2\pi)}{\sqrt{2\beta}} \right) \ \right) \ \right]. \label{3-11} 
\end{align}

\subsection{\normalsize Cylinder ($g=1$) contribution}
Next, we focus on the most dominant contribution from geometry including an observable baby universe, which corresponds to the cylinder geometry ($g=1$). Similarly to the disk geometry, we perform the calculation with the assumption $E,T \gg 1$. We extract the contribution of the cylinder geometry from (\ref{2-23}):\\
\begin{align}
& \langle \mathcal{O}(x_1) \mathcal{O}(x_2)\rangle_{\text{baby}}^{\chi=-1}  = e^{-S_0} \int^{\infty}_{-\infty}e^{\ell}d\ell e^{-\Delta \ell}  \int d\tau db \int^{\infty}_{0} dE_1dE_2 \ e^{-\frac{\beta}{2}(E_1+E_2) } \rho_{\text{Tr}}(E_1,b)\rho_{\text{Tr}}(E_2,b) \psi_{E_1}(\ell)\psi_{E_2}(\ell). \label{3-12} 
\end{align}  
\\
As in the previous subsection, by setting $E_1$ and $E_2$ as in (\ref{3-4}) and using (\ref{3-6}), we obtain\\
\begin{align}
\langle \mathcal{O}(x_1) \mathcal{O}(x_2)&\rangle_{\text{baby}}^{\chi=-1}  = e^{-S_0} \int^{\infty}_{-\log E}e^{\ell}d\ell e^{-\Delta \ell}  \int d\tau db \int^{\infty}_{-\infty} d\omega \ e^{-iT\omega} \rho_{\text{Tr}}(E_1,b)\rho_{\text{Tr}}(E_1,b) \psi_{E_1}(\ell)\psi_{E_2}(\ell) \nonumber \\
& \approx \ \frac{e^{-S_0-2\pi \sqrt{2E}}}{\pi (2E)^{3/2}} \ \int^{\infty}_{-\log E} d\ell e^{-\Delta \ell}  \int d\tau db \int^{\infty}_{-\infty} d\omega \ e^{-iT\omega} \nonumber \\
& \times \left(\ e^{2i\sqrt{2E}L-\frac{i\pi}{2}} +e^{\frac{i\omega}{\sqrt{2E}}L} + e^{-\frac{i\omega}{\sqrt{2E}}L} +  e^{-2i\sqrt{2E}L+\frac{i\pi}{2}}\ \right) \left(\ e^{2i\sqrt{2E}b} +e^{\frac{i\omega}{\sqrt{2E}}b} + e^{-\frac{i\omega}{\sqrt{2E}}b} +  e^{-2i\sqrt{2E}b}\ \right). \label{3-13} 
\end{align}
\\
Under the assumption $E \gg 1$, the four rapidly oscillating terms containing $e^{\pm i\sqrt{2E}L}$ or $e^{\pm i\sqrt{2E}b}$ can be neglected. In this case, four delta functions are generated from the $\omega$ integral as follows:\\
\begin{align}
\langle \mathcal{O}(x_1) \mathcal{O}(x_2)\rangle_{\text{baby}}^{\chi=-1}  & = \frac{2e^{-S_0-2\pi \sqrt{2E}}}{(2E)^{3/2}} \ \int^{\infty}_{-\log E} d\ell e^{-\Delta \ell}\int d\tau db \nonumber \\
&\times \left[ \delta \bigg(T-\frac{(L+b)}{\sqrt{2E}} \bigg) + \delta \bigg(T-\frac{(L-b)}{\sqrt{2E}} \bigg) + \delta \bigg(T+\frac{(L-b)}{\sqrt{2E}} \bigg)  + \delta \bigg(T+ \frac{(L+b)}{\sqrt{2E}} \bigg)  \right] \nonumber \\
& = \frac{e^{-S_0-2\pi \sqrt{2E}}}{E} \int^{\infty}_{-\log E} d\ell e^{-\Delta \ell}\int db d\tau \ \delta(\ell \pm \ell_{T} \pm b) . \label{3-14} \\ \nonumber
\end{align}
Since $T \gg 1$ and $b>0$, $\delta(\ell + \ell_T + b)$ can be dropped and the three remaining delta functions are of interest. The three contributions represent different kinds of geometry that arise from the strip approximation\footnote{According to the Gauss-Bonnet theorem, as the boundary is extended infinitely while the area of the surface is kept, the shape can be approximated as a thin strip.} \cite{Firewall-1}. In \cite{Firewall-1}, the moduli integral was performed within the restricted region satisfying the ``no-shortcut'' condition \footnote{The strip approximation was generalized to  $g \geq 2$ in \cite{Firewall-2}. The integration range of moduli parameters ${b_i, \tau_i}$ is determined from the no-shortcut condition in the same way as in the case $g=1$.}.

\begin{figure}[H] 
\centering 
\includegraphics[width=150mm]{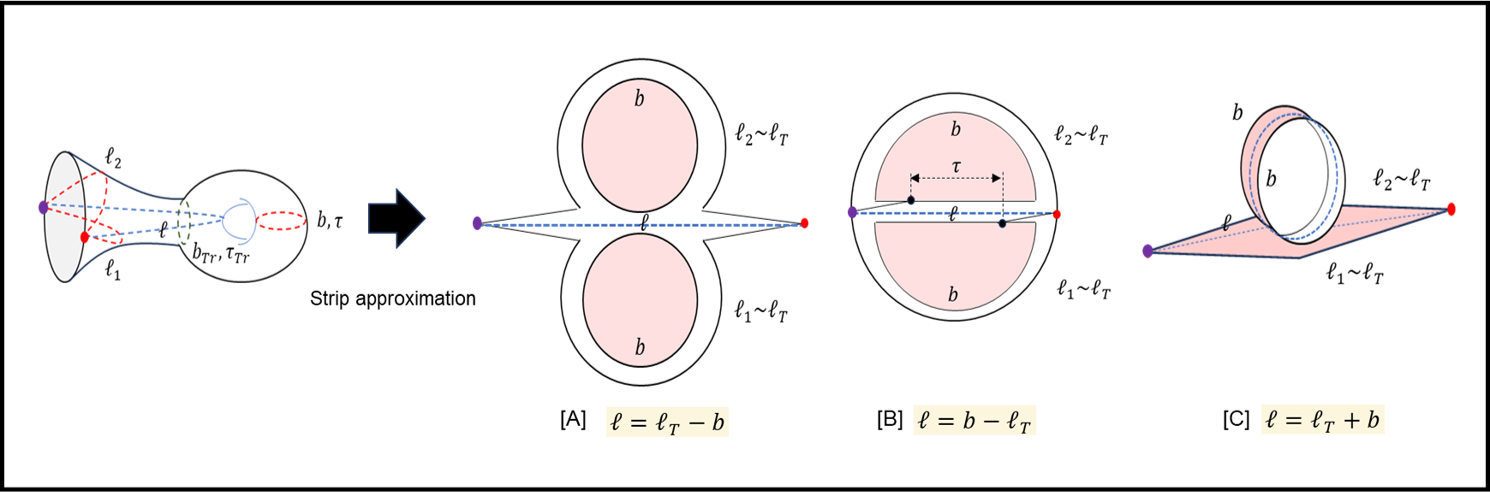} 
\caption{\small Depending on the value of the twist parameter, there may be a shorter geodesic connecting the boundary operators than the one with length $\ell$ shown by the dotted line. In such cases, the regions of the moduli $b$ and $\tau$ are restricted.}
\label{fig3-2(strip)} 
\end{figure}

For $\delta(\ell - \ell_T + b)$ in (\ref{3-14}) (see Fig.\ref{fig3-2(strip)} [A]), the no-shortcut condition imposes no restrictions on the moduli parameters. The integration region is defined as $0 < \tau < b$. Also, taking into account the integration range of $b$, that of $\ell$ becomes $-\log E<\ell<\ell_T$. The upper bound on $\ell$ can also be understood from the fact that there is always a geodesic of length $\sim \ell_T$ connecting the two operators as shown in Fig.\ref{fig3-2(strip)}[A]. Thus, the correlation function is given by:\\
\begin{align}
\langle \mathcal{O}(x_1) \mathcal{O}(x_2)\rangle_{\text{baby}}^{\chi=-1}  &\supset \frac{e^{-S_0-2\pi \sqrt{2E}}}{E} \int^{\ell_T}_{-\log E} d\ell e^{-\Delta \ell}\int^{\ell_T+\log E}_{0} db \int^{b}_{0}d\tau \delta(\ell-\ell_T+b) \nonumber \\
& = \frac{e^{-S_0-2\pi \sqrt{2E}}}{E} \left[\ \left( \ell_T + \log E - \frac{1}{\Delta} \right)\frac{E^{\Delta}}{\Delta} + \frac{e^{-\Delta \ell_T}}{\Delta^{2}}  \ \right]. \label{3-15} \\ \nonumber
\end{align}
On the other hand, $\delta(\ell + \ell_T - b)$ corresponds to the geometry shown in Fig.\ref{fig3-2(strip)} [B], where the no-shortcut conditions impose $\tau > \ell$ and $b - \tau > \ell$. Thus the integral region of $\tau$ becomes $b - \ell_T < \tau < \ell_T$. That said, this contribution is the same as in (\ref{3-15}):\\ 
\begin{align}
\langle \mathcal{O}(x_1) \mathcal{O}(x_2)\rangle_{\text{baby}}^{\chi=-1}  &\supset \frac{e^{-S_0-2\pi \sqrt{2E}}}{E} \int^{\ell_T}_{-\log E} d\ell e^{-\Delta \ell}\int^{2\ell_T}_{\ell_T-\log E} db \int^{\ell_T}_{b-\ell_T}d\tau \delta(\ell+\ell_T-b) \nonumber \\
& = \frac{e^{-S_0-2\pi \sqrt{2E}}}{E} \left[\ \left( \ell_T + \log E- \frac{1}{\Delta} \right)\frac{E^{\Delta}}{\Delta} + \frac{e^{-\Delta \ell_T}}{\Delta^{2}}  \ \right]. \label{3-16} 
\end{align}
\noindent \quad The correlation function contains terms with a factor $e^{-\Delta \ell_T}$ that decays exponentially in time. But, unlike the disk contribution, it also has a term that grows linearly in $T$. Taking the $\tau$-scaling limit ($T, S_0 \to \infty$) discussed in the following subsection, (\ref{3-16}) becomes \\
\begin{align}
&\frac{e^{-S_0-2\pi \sqrt{2E}}}{E} \left[\ \left( \ell_T + \log E- \frac{1}{\Delta} \right)\frac{E^{\Delta}}{\Delta} + \frac{e^{-\Delta \ell_T}}{\Delta^{2}}  \ \right] \ \underset{T,S_0 \to \infty} \rightarrow \ \frac{\sqrt{2}e^{-S_0-2\pi \sqrt{2E}}}{\Delta} TE^{\Delta-\frac{1}{2}}. \label{3-17} \\ \nonumber
\end{align}
Similarly to the disk contribution, by applying the Laplace transform to (\ref{3-17}) one obtains \\
\begin{align}
\langle \mathcal{O}(x_1) \mathcal{O}(x_2)\rangle_{\text{baby}}^{\chi=-1(\beta)}  &\approx \sqrt{2}e^{-S_0}T \int^{\infty}_{0} \frac{dE}{\sqrt{E}}e^{-\beta E}e^{-2\pi \sqrt{2E}} \nonumber \\
& = \frac{\sqrt{2\pi}Te^{-S_0+\frac{2\pi^{2}}{\beta}}}{\sqrt{\beta}} \left[1-\text{Erf}\left( \sqrt{\frac{2\pi^{2}}{\beta}}\right) \right]. \label{3-18} \\ \nonumber
\end{align}
\quad Finally, we discuss the strip geometry corresponding to $\delta(\ell - \ell_{T} - b)$ (see Fig.\ref{fig3-2(strip)} [C]). According to \cite{Firewall-1}, the moduli space for this geometry does not contain any regions that satisfy the no-shortcut condition. Indeed, for any choice of $b$ or $\tau$, the geodesic of length $\ell$ ($\ell$-geodesic) is never shorter than the $\ell_1$-geodesic or $\ell_2$-geodesic because of the excursion around the $b$ cycle. Thus, the shorter of the $\ell_1$-geodesic or $\ell_2$-geodesic should be chosen as the correct spatial slice connecting the two boundary operators. Therefore this situation does not include an observable baby universe. So the contribution from this strip geometry should rather be added to (\ref{3-10}). Thus, this strip geometry is omitted from our discussion here, as it does not affect the behavior of correlation function at late times.

So far, we have calculated the correlation function using the approach of \cite{Firewall-1}. Now, for comparison, let's calculate the correlation function by summing over all geodesics connecting boundary operators, as in \cite{Late,Volume}. 

Once again, we concentrate on the leading contributions to the correlators from the geometries with and without an observable baby universe. With regard to the leading ($g=0$) contribution from geometries without baby universes, the analysis goes in precisely the same way as before and leads to (\ref{3-11}). As for the $g=1$ geometries with a baby universe, the analysis is the same up to (\ref{3-14}). But, after decomposing it into three different types of strip geometries, we integrate over $b$, $\tau$ and $\ell$ (for the geometry shown in [A] and [B] of Fig.\ref{fig3-2(strip)}) ${\it without}$ no-shortcut conditions. As a result, the integral over the strip geometries satisfying $\ell=\ell_T-b$ remains the same as (\ref{3-15}), but the integral over those satisfying $\ell=b-\ell_T$ changes to: \\
\begin{align}
\langle \mathcal{O}(x_1) \mathcal{O}(x_2)\rangle_{\text{baby}}^{\chi=-1}  &\supset \frac{e^{-S_0-2\pi \sqrt{2E}}}{E} \int^{\infty}_{-\log E} d\ell e^{-\Delta \ell}\int^{\infty}_{\ell_T-\log E} db \int^{b}_{0}d\tau \delta(\ell+\ell_T-b) \nonumber \\
& = \frac{e^{-S_0-2\pi \sqrt{2E}}}{E} \ \left( \ell_T -\log E + \frac{1}{\Delta} \right)\frac{E^{\Delta}}{\Delta}\ \underset{T,S_0 \to \infty} \rightarrow \ \frac{\sqrt{2}e^{-S_0-2\pi \sqrt{2E}}}{\Delta} TE^{\Delta-\frac{1}{2}}  . \label{3-19} \\ \nonumber
\end{align}
This result matches with the leading term in (\ref{3-16}) at $T \gg 1$. It also shows a linear growth in $T$ at late times. Note that the mismatch between (\ref{3-16}) and (\ref{3-19}) is of order $\sim e^{-\Delta \ell_T}$ which is suppressed at late times. As explained just before (\ref{3-15}), it can be understood as arising from the second shortest geodesic of length $\ell_T$.

By the integration over strip geometry satisfying $\ell = \ell_T + b$, we get \\
\begin{align}
\langle \mathcal{O}(x_1) \mathcal{O}(x_2)\rangle_{\text{baby}}^{\chi=-1}  &\supset \frac{e^{-S_0-2\pi \sqrt{2E}}}{E} \int^{\infty}_{\ell_T} d\ell e^{-\Delta \ell}\int^{\infty}_{0} db \int^{b}_{0}d\tau \delta(\ell-\ell_T-b) \nonumber \\
& = \frac{e^{-S_0-2\pi \sqrt{2E}}}{\Delta^2 E} e^{-\Delta \ell_T} \ \underset{T,S_0 \to \infty} \rightarrow 0. \label{3-20} \\ \nonumber
\end{align}
For this strip geometry, the length $b_{\text{Tr}}$ of the closed geodesic which is homologous to the asymptotic AdS boundary is small. According to \cite{Firewall-1}, the contribution to the two-point function from the surface with small $b_{\text{Tr}}$ is actually negative, but we obtain a small positive value. Anyway this contribution should be considered as a small correction to the disk as explained in the previous paragraph.

We have seen that the difference between the two approaches is exponentially small at late times. But it becomes noticeable as $\Delta \to 0$. Indeed, by taking the limit of  (\ref{3-16}) and (\ref{3-19}) one finds \\
\begin{align}
\lim_{\Delta \to 0} \text{(\ref{3-16})} &= \frac{e^{-S_0-2\pi\sqrt{2E}}}{E} \left( \ \frac{\ell_T^2}{2}+\ell_T + \frac{1}{2}\left(\log E \right)^2 \ \right), \nonumber \\
\lim_{\Delta \to 0} \text{(\ref{3-19})} &= \lim_{\Delta \to 0} \left[ \frac{1}{\Delta^2} + \frac{\ell_T}{\Delta} +\left(\ell_T-\log E +\frac{1}{2} \right)\log E +O(\Delta) \right] \to \infty. \label{3-20-2} \\ \nonumber
\end{align}
This implies that the correlation function defined by the no-shortcut condition is finite in the limit $\Delta \to 0$ and agrees with the partition function by definition, but the one defined by summing over all the geodesics diverges.

\subsection{\normalsize Behavior at late times}
Let's next study the correlator at late times by taking the $\tau$-scaling limit \cite{OS}:\\
\begin{align}
T \to \infty, \quad e^{S_0} \to \infty, \quad Te^{-S_0}:\text{fixed}. \label{3-21}  
\end{align}
\\
Since $T$ and the energy difference $\omega$ are conjugate to each other, the above limit is equivalent to :\\
\begin{align}
\omega \to 0, \quad e^{S_0} \to \infty, \quad \omega e^{S_0}:\text{fixed}. \label{3-22}  
\end{align}
\\
Also, since JT gravity has a matrix model dual, its spectral density correlator $\langle \rho(E_1)\rho(E_2) \rangle$ for small $E_1-E_2$ should exhibit the universal behavior  \cite{SSS,sinkernel-1, sinkernel-2}:\\
\begin{align}
\langle \rho(E_1)\rho(E_2) \rangle \approx e^{2S_0} \hat{\rho}_0(E)^{2} + e^{S_0} \delta(\omega) \hat{\rho}_0(E) - \frac{\sin^{2}(\pi e^{S_0} \hat{\rho}_{0}(E)\omega)}{\pi^{2} \omega^{2}}, \label{3-23} \\ \nonumber 
\end{align}
where $E_1$, $E_2$ are related to $E$, $\omega$ as in (\ref{3-4}).

In this limit, the non-perturbative correction terms drop out, and it is known that the main physical information can be captured by the perturbative sum. The formula (\ref{3-23}) contains all-genus information, although the contribution from each genus has been somewhat simplified because of the $\tau$-scaling limit \cite{SSS,plateau1}.

Now, using (\ref{3-23}) to express the two-point correlation function with the energy fixed, we obtain \\
\begin{align}
\langle \mathcal{O}(x_1) \mathcal{O}(x_2)\rangle_{\text{universal}}  &\approx e^{-S_0} \int^{\infty}_{-\log E} e^{\ell} e^{-\Delta \ell} \ d\ell \int^{\infty}_{-\infty} d\omega e^{-iT\omega} \psi_{E+\frac{\omega}{2}}(\ell)\psi_{E-\frac{\omega}{2}}(\ell) \nonumber \\
& \times \ \left( \ e^{2S_0} \hat{\rho}_0(E)^{2} + e^{S_0} \delta(\omega) \hat{\rho}_0(E) - \frac{\sin^{2}(\pi e^{S_0} \hat{\rho}_{0}(E)\omega)}{\pi^{2} \omega^{2}} \ \right). \label{3-24} \\ \nonumber
\end{align}
The first term $\ e^{2S_0} \hat{\rho}_0(E)^{2}$ in the parenthesis reproduces the disk contribution (\ref{3-10}), as expected:\\
\begin{align}
e^{S_0} \int^{\infty}_{-\log E} e^{\ell} e^{-\Delta \ell} \ d\ell \int^{\infty}_{-\infty} d\omega e^{-iT\omega} \psi_{E+\frac{\omega}{2}}(\ell)\psi_{E-\frac{\omega}{2}}(\ell) \hat{\rho}_0(E)^{2}= \frac{e^{S_0+2\pi\sqrt{2E}}}{2\pi^2} e^{-\Delta \ell_T}. \label{3-25} \\ \nonumber
\end{align}
The second term $e^{S_0} \delta(\omega) \hat{\rho}_0(E)$ corresponds to the contact term. By using (\ref{3-6}) and dropping the terms that oscillate rapidly for $E \gg 1$, we obtain \\
\begin{align}
&\ \int^{\infty}_{-\log E} e^{\ell} e^{-\Delta \ell} \ d\ell \int^{\infty}_{-\infty} d\omega \ e^{-iT\omega} \delta(\omega)\psi_{E+\frac{\omega}{2}}(\ell)\psi_{E-\frac{\omega}{2}}(\ell) \hat{\rho}_0(E) \nonumber \\
& \approx \frac{\sqrt{2}}{\pi \sqrt{E}}\ \int^{\infty}_{-\log E}  e^{-\Delta \ell} \ d\ell \int^{\infty}_{-\infty} d\omega \left( \ e^{-i\left( T-\frac{L}{\sqrt{2E}} \right)\omega} + e^{-i\left( T+\frac{L}{\sqrt{2E}} \right)\omega} \ \right) = \frac{\sqrt{2}}{\Delta \pi} E^{\Delta-\frac{1}{2}}. \label{3-26} 
\end{align}
\\
By Laplace transforming the above result, we obtain\\
\begin{align}
\frac{\sqrt{2}}{\Delta \pi}\ \int^{\infty}_{0} dE E^{\Delta-\frac{1}{2}}e^{-\beta E} = \frac{\sqrt{2}\Gamma(\Delta+\frac{1}{2})}{\Delta \pi \beta^{\Delta+\frac{1}{2}}}. \label{3-27} \\ \nonumber
\end{align}
\quad Let's now turn to the third term $-\frac{\sin^{2}(\pi e^{S_0} \hat{\rho}_{0}(E)\omega)}{\pi^{2} \omega^{2}}$. Applying the same approximations as before, we obtain \\
\begin{align}
& \frac{e^{-S_0-2\pi\sqrt{2E}}}{\pi\sqrt{2E}}\ \int^{\infty}_{-\log E}  e^{-\Delta \ell} \ d\ell \int^{\infty}_{-\infty} \frac{d\omega}{\omega^{2}} \nonumber \\
&  \times \ \Biggl( \ e^{-i\left( T-\frac{L}{\sqrt{2E}}-2\pi e^{S_0}\hat{\rho}_0(E) \right)\omega} + e^{-i\left( T-\frac{L}{\sqrt{2E}}+2\pi e^{S_0} \hat{\rho}_0(E) \right)\omega} -2e^{-i\left( T-\frac{L}{\sqrt{2E}}\right)\omega} \nonumber \\
& + \ e^{-i\left( T+\frac{L}{\sqrt{2E}}-2\pi e^{S_0} \hat{\rho}_0(E) \right)\omega} + e^{-i\left( T+\frac{L}{\sqrt{2E}}+2\pi e^{S_0} \hat{\rho}_0(E) \right)\omega} -2e^{-i\left( T+\frac{L}{\sqrt{2E}}\right)\omega } \ \Biggr). \label{3-28} \\ \nonumber
\end{align}
The integral over $\omega$ can be evaluated by using\\
\begin{align}
\int^{\infty}_{-\infty} \frac{d\omega}{\omega^{2}}\ e^{-iT\omega} = -\pi |T|. \label{3-29} \\ \nonumber
\end{align}
We are then left with an integral over $\ell$, which is tedious but straightforward. The end result depends on whether $T$ is bigger or smaller than $T_{H}=2\pi e^{S_0}\hat{\rho}_0(E)$.\\

$\blacksquare$ $T<T_{H}$\\
\begin{align}
\text{(\ref{3-28})} &= \ \frac{2\sqrt{2} T E^{\Delta-\frac{1}{2}}}{\Delta}e^{-S_0-2\pi\sqrt{2E}}- \frac{\sqrt{2}E^{\Delta-\frac{1}{2}}}{\pi \Delta} \nonumber \\
& \hspace{30pt}- \frac{2e^{-S_0}}{E\Delta^{2}} \Bigl[ e^{-2\pi \sqrt{2E}(1+\Delta e^{S_0} \hat{\rho}_0(E)) } \cosh(\sqrt{2E} \Delta T) - e^{-\sqrt{2E}(2\pi + \Delta T)} \Bigr] (2E)^{\Delta}e^{-2\Delta}, \label{3-30} 
\end{align}
\\

$\blacksquare$ $T>T_{H}$\\
\begin{align}
\text{(\ref{3-28})} &= - \frac{4e^{-S_0}}{E\Delta^{2}} e^{-\sqrt{2E}(2\pi+\Delta T)}\ \sinh^{2} \left( \  \pi \Delta \sqrt{2E} e^{S_0} \hat{\rho}_0(E) \ \right)(2E)^{\Delta}e^{-2\Delta}.\hspace{80pt} \label{3-31} \\ \nonumber
\end{align}
Thus the correlation function is obtained by adding (\ref{3-25}) and (\ref{3-26}) to (\ref{3-30}) and (\ref{3-31}):\\

$\blacksquare$ $T<T_{H}$\\
\begin{align}
\langle \mathcal{O}(x_1) \mathcal{O}(x_2)\rangle_{\text{universal}}  &\approx \frac{e^{S_0+2\pi\sqrt{2E}}}{2\pi^2} e^{-\Delta \ell_T} + \ \frac{2\sqrt{2} T E^{\Delta-\frac{1}{2}}}{\Delta}e^{-S_0-2\pi\sqrt{2E}} \nonumber \\
& - \frac{4e^{-S_0}}{\Delta^{2}} \Bigl[ e^{-2\pi \sqrt{2E}(1+\Delta e^{S_0} \hat{\rho}_0(E)) } \cosh(\sqrt{2E} \Delta T) - e^{-\sqrt{2E}(2\pi + \Delta T)} \Bigr] (2E)^{\Delta-1}e^{-2\Delta},  \label{3-32} \\ \nonumber
\end{align}

$\blacksquare$ $T>T_{H}$\\
\begin{align}
\langle \mathcal{O}(x_1) \mathcal{O}(x_2)\rangle_{\text{universal}}  &\approx \frac{e^{S_0+2\pi\sqrt{2E}}}{2\pi^2} e^{-\Delta \ell_T} + \frac{\sqrt{2}}{\Delta \pi} E^{\Delta-\frac{1}{2}}  \nonumber \\
&- \frac{8e^{-S_0}}{\Delta^{2}} e^{-\sqrt{2E}(2\pi+\Delta T)}\ \sinh^{2} \left( \  \pi \Delta \sqrt{2E} e^{S_0} \hat{\rho}_0(E) \ \right)(2E)^{\Delta-1}e^{-2\Delta}. \hspace{60pt} \label{3-33} 
\end{align}
\\
Finally, in the limit $T,S_0 \to \infty$ we are only left with: \\
\begin{align}
\langle \mathcal{O}(x_1) \mathcal{O}(x_2)\rangle_{\text{universal}} \ \underset{T,S_0 \to \infty} \rightarrow 
\begin{cases}
\quad \displaystyle{\frac{2\sqrt{2}T}{\Delta} e^{-S_0-2\pi \sqrt{2E}}\ E^{\Delta-\frac{1}{2}}}  & (T<T_H) \\
\\
\quad \displaystyle{\frac{\sqrt{2}}{\Delta \pi} E^{\Delta-\frac{1}{2}}} & (T>T_H). 
\end{cases} \label{3-34} \\ \nonumber 
\end{align}
For $ T < T_H$, the results match those of (\ref{3-17}) and (\ref{3-19}), except for a factor of 2 in the coefficients (this discrepancy in the coefficients is resolved by combining the relevant contributions from (\ref{3-15}) and (\ref{3-16})). This extracts the contribution from $g=1$ in (\ref{3-23}), giving the ``ramp''. We see that there is agreement between the three methods when focusing on the late times. Also, for $T>T_H$ we obtain a ``plateau'' which was not observed in the previous two results. Overall, our results are consistent with those for the SFF, with no permanent decrease in correlations. 

\section{The length of the ERB and the size of baby universes at late times}\label{ERB}
Finally, in this section, we consider ERB as another clue to understanding the behavior of baby universes.

According to \cite{Late}, the two are related in that after an ERB that has been growing for a long time stops growing, a part of the saturated ERB is cut off and a baby universe is emitted. Also in this process, the probability of baby universe emission rises as $\langle \ell \rangle$ increases, and the size of the baby universe will be very close to the length of the ERB at the time of emission. Then, as $T$ approaches $T_H$, the growth of the ERB length slows down, and the baby universes emission balances with the growth of the ERB.

Based on this argument, in this section we aim to study the state of the ERB and baby universes at late times. First, following \cite{Volume,TTbar}, we review the behavior of the ERB in the $\tau$-scaling limit and find that it stops growing around $T=T_H$. Next, we use this result to calculate the size of the ERB and baby universes at late times $T \sim T_H$, using the geometry with one observable baby universe (ramp contribution to the correlation function).
\subsection{\normalsize The length of the ERB in the $\tau$-scaling limit}
In JT gravity, the volume inside a two-sided black hole is estimated as the length $\ell$ of the ERB connecting the two sides. For a simply connected surface with trivial topology, the choice of ERB is unique. But, for surfaces with higher genus, there are infinitely many geodesics connecting two sides, making the definition of $\ell$ ambiguous. As a concrete definition of $\langle \ell \rangle$, we adopt the one proposed in \cite{Volume}:\\
\begin{align}
\langle \ell \rangle = \lim_{\Delta \to 0} \bigg\langle \sum_{\gamma} \ \ell_{\gamma} e^{-\Delta \ell_{\gamma}} \bigg\rangle. \label{4-1}  \\ \nonumber
\end{align}
This definition is well-defined on surfaces with any topology. It minimizes the backreaction on the metric and ensures the analytic continuation between Euclidean and Lorentzian geometries \cite{susskind, Volume, Volume-2}. Here we take the sum over all non-self-intersecting geodesics, and $\Delta$ acts as a regularization parameter.

The correlation function of boundary operators is expressed as a sum of geodesics as in (\ref{2-10}). Thus, $\langle \ell \rangle$ can be obtained by differentiating (\ref{2-10}) with respect to $\Delta$ and taking the limit $\Delta \to 0$:\\
\begin{align}
\langle \ell \rangle = -\lim_{\Delta \to 0} \ \frac{\partial \langle \mathcal{O}(x_1)\mathcal{O}(x_2) \rangle}{\partial \Delta}. \label{4-2} \\ \nonumber
\end{align}
Now, recalling (\ref{2-5}) and (\ref{3-23}), the two-point correlation function is given by:\\
\begin{align}
\langle \mathcal{O}(x_1)\mathcal{O}(x_2) \rangle_{\text{universal}} \approx e^{-S_0}\int^{\infty}_{-\infty} d\omega e^{-iT\omega}\bigg[\ e^{2S_0} \hat{\rho}_0(E)^{2} + e^{S_0} \delta(\omega) \hat{\rho}_0(E) - \frac{\sin^{2}(\pi e^{S_0} \hat{\rho}_{0}(E)\omega)}{\pi^{2} \omega^{2}} \ \bigg]\ |\mathcal{O}_{E_1,E_2}|^{2}. \label{4-3} \\ \nonumber
\end{align}
The derivative with respect to $\Delta$ acts only on $|\mathcal{O}_{E_1,E_2}|^{2}$. By calculating this and taking the limit $\Delta \to 0$, we obtain (see Appendix \ref{Appendix-ERB-1}):\\
\begin{align}
\frac{\partial}{\partial \Delta}|\mathcal{O}_{E_1,E_2}|^2 &= \frac{\delta(\omega)}{4\hat{\rho}_0(E)} \big\{ \psi(2i\sqrt{2E}) + \psi(-2i\sqrt{2E}) -2\log2 \big\} + \frac{1}{8\pi^2 \hat{\rho}_0(E)\hat{\rho}_0(\omega)\omega}, \label{4-4} \\ \nonumber
\end{align}
where the following formulae have been used \\
\begin{align}
\psi(x)=\frac{d \log \Gamma(x)}{dx}=\frac{\Gamma'(x)}{\Gamma(x)}, \quad \hat{\rho}_{0}(\omega)=\frac{1}{2\pi^{2}}\sinh \left( \frac{\pi \omega}{\sqrt{2E}} \right). \label{4-5} \\ \nonumber
\end{align}
Thus, $\langle \ell(T) \rangle$ is given by:\\
\begin{align}
\langle \ell(T) \rangle &= \text{const} \ - \ \frac{e^{-S_0}}{8\pi^{2}\hat{\rho}_0(E)} \int^{\infty}_{-\infty} \frac{d\omega}{\omega \hat{\rho}_{0}(\omega)}e^{-iT\omega} \ \bigg[\ e^{2S_0} \hat{\rho}_0(E)^{2}  - \frac{\sin^{2}(\pi e^{S_0} \hat{\rho}_{0}(E)\omega)}{\pi^{2} \omega^{2}} \ \bigg],  \label{4-6} \\ \nonumber
\end{align}
where the $T$-independent terms, some of which are divergent, are denoted collectively as const\footnote{Substituting (\ref{4-4}) into (\ref{4-3}) one obtains terms $\sim e^{-iT\omega}\delta(\omega)$ and $\sim e^{-iT\omega}\delta^2(\omega)$ in the integrand. The ``const'' in (\ref{4-6}) arises from integrating these terms over $\omega$, but the integral of terms $\sim \delta^{2}(\omega)$ is apparently divergent. Regularization of this divergence requires tracing back to the origin of this delta function. Since this goes beyond the scope of this paper, we will not delve into it here.}. As in \cite{Volume}, we simply discard this divergence and study the finite and T-dependent part $\langle \ell(T)-\ell(0) \rangle$.

\begin{figure}[H] 
\centering 
\includegraphics[width=61mm]{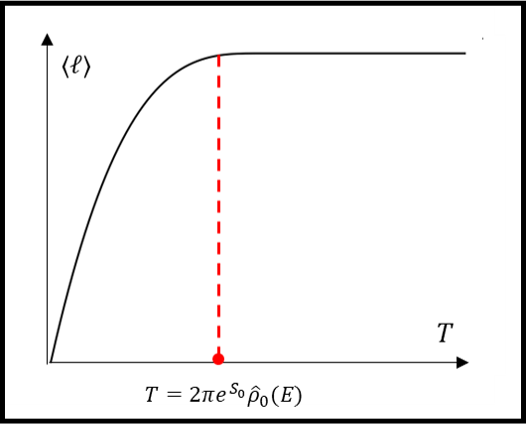} 
\caption{\small Time evolution of the (regularized) ERB length $\langle \ell \rangle$. The growth stops at $T=T_H$.} 
\label{fig4-1(sinkernel-1)} 
\end{figure}

We evaluate the integral over $\omega$ in (\ref{4-6}) using (\ref{3-28}) and the following formula:\\
\begin{align}
\int^{\infty}_{-\infty} \frac{d\omega}{\omega^{4}}\ e^{-iT\omega} = \frac{\pi |T|^{3}}{6}.  \label{4-7} \\ \nonumber
\end{align}
As a result, we obtain (see Fig.\ref{fig4-1(sinkernel-1)})\\ 
\begin{align}
\langle \ell(T) \rangle=
\begin{cases}
\quad \displaystyle{\text{const}\ - \ \frac{e^{2S_0}\pi (2E)^{1/2} \hat{\rho}^2_{0}(E)}{6}\ \left(1-\frac{T}{2\pi e^{S_0} \hat{\rho}_0(E)} \right)^{3}}  & (T<T_H) \\ 
\\
\quad \displaystyle{\text{const}} & (T>T_H) .
\end{cases} \label{4-8} \\ \nonumber
\end{align}
From this result, it is evident that $\langle \ell(T) \rangle$ changes its behavior at $T = T_H$ \cite{Miyaji}. For $T<T_H$, it grows as a cubic function in $T$, but beyond that, no growth is observed and it remains constant. Thus $T = T_H$ is an important critical time at which emission of the baby universe and the ERB balance.
\subsection{\normalsize The length of the ERB in geometry including one observable baby universe}
Next we evaluate $\langle \ell(T) \rangle$ by applying the method of \cite{Late,Volume} and taking the $g=1$ geometry which includes one observable baby universe.

We begin with the following term in the correlator with fixed energy:\\
\begin{align}
\langle \mathcal{O}(x_1)\mathcal{O}(x_2) \rangle_{\text{baby}}^{\chi=-1} &=  e^{-S_0} \ \int^{\infty}_{0} bdb  \int^{\infty}_{-\infty} d\omega e^{-iT\omega} \ \rho_{\text{Tr}}(E_1,b) \rho_{\text{Tr}}(E_2,b)  |\mathcal{O}_{E_1,E_2}|^{2}. \label{4-9} \\ \nonumber
\end{align}
By using (\ref{4-4}) and assuming $E \gg 1 $, we discard the rapidly oscillating terms, and $\langle \ell(T) \rangle$ is given by:\\
\begin{align}
\langle \ell(T) \rangle_{\chi=-1} &= \text{const} - \ \frac{e^{-S_0}}{8\pi^2 \hat{\rho}_0(E)} \ \int^{\infty}_{0} bdb \int^{\infty}_{-\infty} \frac{d\omega}{\omega \hat{\rho}_0(\omega)} e^{-iT\omega} \ \rho_{\text{Tr}}(E_1,b) \rho_{\text{Tr}}(E_2,b) \nonumber \\
&\sim  \text{const} - \ \frac{e^{-S_0}}{16\pi^3 \hat{\rho}_0(E)(2E)^{1/2}} \ \int^{\infty}_{0} bdb \int^{\infty}_{-\infty} \frac{d\omega}{\omega^{2}} \Big( e^{-i\left(T-\frac{b}{\sqrt{2E}}\right)\omega } + e^{-i\left(T+\frac{b}{\sqrt{2E}}\right)\omega } \Big). \label{4-10} \\ \nonumber
\end{align}
By integrating with respect to $b$ and $\omega$ we obtain\\
\begin{align}
\langle \ell(T) \rangle_{\chi=-1} =  \text{const} \ + \ \frac{e^{-S_0}(2E)^{1/2}}{48\pi^2 \hat{\rho}_0(E)}\ T^{3}. \label{4-11} \\ \nonumber
\end{align}
Due to the contribution from geometry including one observable baby universe, $\langle \ell(T) \rangle$ grows as a cubic function in $T$.

Next, let's compare the result (\ref{4-11}) with the solution in the $\tau$-scaling limit. The geometry that gives us (\ref{4-11}) is a disk with one handle, so we can compare it with the term containing $e^{-S_0}$ in (\ref{4-3}). Thus, we rewrite the sine kernel as follows:\\
\\ 
\begin{align}
\langle \mathcal{O}(x_1)\mathcal{O}(x_2) \rangle_{\text{universal}} &= e^{-S_0}\int^{\infty}_{-\infty} d\omega e^{-iT\omega}\bigg[\ e^{2S_0} \hat{\rho}_0(E)^{2} + e^{S_0} \delta(\omega) \hat{\rho}_0(E) \underset{\text{factors of $e^{-S_0}$}}{\underline{- \frac{1}{2\pi^2 \omega^2}}} + \frac{\cos(2\pi e^{S_0} \hat{\rho}_{0}(E)\omega)}{2\pi^{2} \omega^{2}} \ \bigg]\ |\mathcal{O}_{E_1,E_2}|^{2}. \label{4-12} 
\end{align}
\\
Extracting the third term in the bracket we obtain\\
\begin{align} 
\langle \mathcal{O}(x_1)\mathcal{O}(x_2) \rangle_{\text{universal}}^{\chi=-1} \ = \ -e^{-S_0}\int^{\infty}_{-\infty} d\omega e^{-iT\omega} \frac{1}{2\pi^2 \omega^2} |\mathcal{O}_{E_1,E_2}|^{2}. \label{4-13} \\ \nonumber
\end{align}
The expectation value $\langle \ell(T) \rangle_{\chi=-1}$ for this correlation function is given by:\\
\begin{align}
\langle \ell(T) \rangle_{\chi=-1} \ &= \ \text{const} \ + \ \frac{e^{-S_0}}{16\pi^4 \hat{\rho}_0(E) } \int^{\infty}_{-\infty} \frac{d\omega}{\omega^3 \hat{\rho}_0(\omega)}\ e^{-iT\omega} \nonumber \\
&= \  \text{const} \ + \ \frac{e^{-S_0}(2E)^{1/2}}{48\pi^2 \hat{\rho}_0(E)}\ T^{3}. \label{4-14} \\ \nonumber
\end{align}
This result is consistent with (\ref{4-11}). The $T^{3}$ term in $\langle \ell(T) \rangle$ comes from geometry including one observable baby universe and contributes significantly to the growth of $\langle \ell(T) \rangle$.
\subsection{\normalsize The expectation value of the size $b$ of the baby universe}
Finally, we evaluate the expectation value of the size of the baby universe, $\langle b(T) \rangle$. Since it is expected to be very close to the ERB length at the time of its emission, it is expected that $\langle b(T) \rangle \sim T^{3}$. We define $\langle b(T) \rangle$ in a somewhat simplistic manner as follows:\\
\begin{align}
\langle b(T) \rangle \equiv \lim_{\Delta \to 0} \langle b \mathcal{O}(x_1) \mathcal{O}(x_2)\rangle_{\chi=-1}. \label{4-15} \\ \nonumber
\end{align}
Then, $\langle b(T) \rangle$ can be expressed as follows:\\
\begin{align}
\langle b(T) \rangle = \lim_{\Delta \to 0} e^{-S_0} \ \int^{\infty}_{-\infty} e^{\ell} d\ell e^{-\Delta \ell} \int bdb \int d\tau \ \int^{\infty}_{-\infty} d\omega e^{-iT\omega} \rho_{\text{Tr}}(E_1,b)\rho_{\text{Tr}}(E_2,b)\psi_{E_1}(\ell)\psi_{E_2}(\ell). \label{4-16} \\ \nonumber
\end{align}
By starting from the expression similar to (\ref{3-14}) and integrating over $\omega$, the following delta functions are obtained\\
\begin{align}
\langle b(T) \rangle &\approx \lim_{\Delta \to 0} \frac{e^{-S_0-2\pi \sqrt{2E}}}{E} \int^{\infty}_{-\log E} d\ell e^{-\Delta \ell}\int bdb \ d\tau \ \bigl( \ \delta(\ell -\ell_T+b) + \delta(\ell +\ell_T-b) \ \bigr), \label{4-17} \\ \nonumber
\end{align}
where we consider only the case containing an observable baby universe. This time, we proceed with the calculations following the method of \cite{Firewall-1}. In this case, particularly for the geometry corresponding to $\delta(\ell+\ell_T-b)$, the no-shortcut condition imposes restrictions on the moduli space:\\
\begin{align}
&\langle b(T) \rangle \approx  \lim_{\Delta \to 0} \frac{e^{-S_0-2\pi \sqrt{2E}}}{E} \ \Big[ e^{-\Delta \ell_T} \ \int^{\ell_T+\log E}_{0} be^{\Delta b}db \int^{b}_{0}d\tau + e^{\Delta \ell_T} \ \int^{2\ell_T}_{\ell_T-\log E} be^{-\Delta b}db \int^{\ell_T}_{\ell_T-b}d\tau \ \Big].  \label{4-18} \\ \nonumber
\end{align}
Thus, $\langle b(T) \rangle$ is given by:\\
\begin{align}
\langle b(T) \rangle &\approx  \lim_{\Delta \to 0} \ \frac{e^{-S_0-2\pi \sqrt{2E}}}{E\Delta}\ \Big\{\ 2\ell_{T}(\ell_T + \log E)E^{\Delta} -\frac{2\ell_T}{\Delta}(E^{\Delta}-e^{-\Delta \ell_T})\Big\} \approx \frac{e^{-S_0}(2E)^{1/2}}{2\pi^{2} \hat{\rho}_0(E)} T^{3}. \label{4-19} \\ \nonumber
\end{align}
Similarly to (\ref{4-11}), we find that the size of the baby universe has $T^{3}$ and grows at the same rate as $\langle \ell(T) \rangle$.

Overall, we find that there is good agreement in the results between the three different methods, even for the sizes of the ERB and the baby universe.
\section{Discussion}
In this paper, we calculated the two-point correlation function in JT gravity, expectation value of the ERB length, and the size of the baby universe using three different approaches. The correlation functions at late times obtained from the three approaches all exhibited a ``ramp'' behavior. Furthermore, we found that one of the three methods can also reproduce the ``plateau'' behavior. These behaviors are consistent with the SFF analysis. Our result indicates that the ramp behavior is attributed to the geometry with one baby universe. We also found that, due to the contribution of this geometry, the ERB length $\langle \ell(T) \rangle$ grows as a cubic function of $T$ at late times $T \lesssim T_H$, but it converges to a constant at $T>T_H$. 

Ultimately, we are interested in extending our perturbative analysis on $\langle \ell(T) \rangle$ and $\langle b(T) \rangle$ to higher genus. In particular, we wish to define and calculate the number and size of observable baby universes when there are multiple of them. For higher genus surfaces, integration over the moduli space is undoubtedly a complicated problem, but it might become simpler by applying the approximation scheme we employed for observables at late times, such as strip approximation. It is indeed remarkable that, despite all kinds of the three approaches were consistent with one another and also with the known late-time behavior of SFF. This fact gives us confidence that the same approximation method works for higher genus correction as well.

\appendix
\section{The behavior of the SFF at late times} \label{AppendixSFF}
Here we review the analysis of the spectral form factor (SFF) in JT gravity \cite{SFF-APP-4}.

The SFF is a useful indicator for measuring the energy level statistics of quantum chaotic systems and has been widely studied in many areas of physics. In particular, in \cite{SFF-APP-1,SFF-APP-2,SFF-APP-3} the SFF for the SYK model was shown to exhibit ``ramp'' and ``plateau'' behavior at late times. The results of SFF analysis are expected to provide clues to resolving the black hole information loss problem.

The SFF is defined as the ensemble average $\left\langle Z(\beta+iT)Z(\beta-iT)\right\rangle$ and is expressed in terms of the eigenvalue density correlation function as follows:\\
\begin{align}
\langle Z(\beta+i T)Z(\beta-i T)\rangle= \int^{\infty}_{0} dE_1 dE_2 \ \langle \rho(E_1)\rho(E_2)\rangle \ e^{-\beta(E_1+E_2)} e^{-i T(E_1-E_2)}. \label{A-1} \\ \nonumber
\end{align}
In JT gravity at large $e^{S_0}$, for small energy separation $\vert E- E'\vert \ll 1$ and sufficiently far away from spectral edges such that the spectral density $\rho_0(E)$ is not wildly varying, the density correlator takes the form \cite{SSS,sinkernel-1,sinkernel-2}: \\
\begin{align}
\left\langle \rho(E_1)\rho(E_2)\right\rangle & \approx
\rho_0(E_1)\rho_0(E_2) -\frac{\sin^2(\pi \rho_0(E_1)(E_1-E_2))}{\pi^2(E_1-E_2)^2} + \rho_0(E_1) \delta(E_1-E_2)  \label{A-2} \\ 
& = \rho_0(E_1) \rho_0(E_2) - \frac{1}{2\pi^2(E_1-E_2)^2} + \frac{\cos(2\pi \rho_0(E_1)(E_1-E_2))}{2\pi^2 (E_1-E_2)^2} + \rho_0(E_1) \delta(E_1-E_2). \label{A-3} \\ \nonumber 
\end{align}
This kind of formula holds for any random matrix potential $V(H)$. This is a direct manifestation of random matrix universality, governing universal level statistics at small energy separations. The right-hand side contains the well-known sine kernel which implements the repulsion of levels in chaotic systems. Indeed, in the limit $E_1\to E_2$, the first two terms in (\ref{A-2}) cancel out, indicating that the energy levels do not want to approach with each other \cite{SSS}.

The first term with a factor of $e^{2S_0}$ is the most dominant, but when the energy difference is very small, $E_1-E_2 \sim e^{-S_0}$, the sine kernel contributes with the same order as the first term. And this contribution then appears as the behavior of SFF at late times which was not expected by classical theory.

Let's use (\ref{A-3}) to analyze the behavior of the SFF at late times.\\
\\
$\blacksquare$ The first term: $\rho_0(E_1)\rho_0(E_2) $

In JT gravity, this arises from two disconnected disks. Each disk gives rise to a partition function:\\
\begin{align}
Z_{D}(\beta+iT) = e^{S_{0}} \int^{\infty}_{0} dE e^{-(\beta+iT)E} \frac{\sinh(2\pi\sqrt{2E})}{2\pi^{2}}= \frac{e^{S_0}}{4\pi^2}\left(\frac{2\pi}{\beta+iT}\right)^{3/2}e^{\frac{2\pi^2}{\beta+iT}}. \label{A-4}\\ \nonumber
\end{align}
This term is the most dominant for $\beta \ll T \lesssim e^{S_0/2}$. This period is called the slope because the SFF exhibits a power law decay in accordance with classical theory:\\ 
\begin{align}
\langle Z_{D}(\beta+iT) Z_{D}(\beta-iT)\rangle \ \sim \  \frac{e^{2S_0}}{2\pi} \frac{1}{T^3} \hspace{20pt} (\beta \ll T \lesssim e^{S_0/2}). \label{A-5}
\end{align}
\\
$\blacksquare$ The second term: $- \frac{1}{2\pi^2(E-E')^2}$

In JT gravity, this corresponds to the double trumpet geometry. The corresponding partition function $Z_{0,2}(\beta_1,\beta_2)$ exhibits a linear growth:\\
\begin{align}
Z_{0,2}(\beta+iT,\beta-iT) = \frac{1}{2\pi} \frac{\sqrt{\beta^2+T^2}}{2\beta} \sim \frac{T}{4\pi \beta}.  \label{A-6} \\ \nonumber
\end{align}
This term is responsible for the growth of SFF which is linear in $T$ during the period $e^{\frac{S_{0}}{2}} \lesssim T \lesssim e^{S_{0}}$, called the ramp:\\
\begin{align}
\langle Z(\beta+iT) Z(\beta-iT)\rangle \sim \frac{T}{4\pi\beta} \hspace{20pt}  (e^{\frac{S_{0}}{2}} \lesssim T \lesssim e^{S_{0}}). \label{A-7} \\ \nonumber
\end{align}
$\blacksquare$ The third term: $\frac{\cos(2\pi \rho_0(E)(E-E'))}{2\pi^2 (E-E')^2}$

This term oscillates rapidly due to $\rho_0(E) \sim e^{S_0} \gg 1$. This represents a non-perturbative correction arising from $\exp(i\# e^{S_0})$.

Through the $\omega$ integration, a term linear in $T$ with negative coefficient is generated at times $T \gtrsim  e^{S_0}$, and stops the linear growth of the SFF after $T \sim  e^{S_0}$. Near $T\sim e^{S_0}$, the SFF is smoothed out due to the remaining Laplace transform in (\ref{A-1}). This causes the spectral form factor at times $T \gtrsim e^{S_0}$ to flatten out and reach the plateau \cite{OS}:\\
\begin{align}
\langle Z(\beta+iT) Z(\beta-iT)\rangle = 
\begin{cases}
\quad \displaystyle{-Z_{D}(2\beta)+\int^{E_T}_{0} dE e^{-2\beta E} \rho_0(E)}  & (T<T_H)  \\
\\
\quad \displaystyle{-\frac{T}{4\pi \beta} \ (1-e^{-\beta E_{T}} )}  & (T>T_H), 
\end{cases} \label{A-8} \\ \nonumber
\end{align}
where $E_T$ satisfies $\hat{\rho}_{0}(E_T)=\frac{e^{-S_0}T}{2\pi}$ and $T_H=2\pi e^{S_0}\hat{\rho}_0(E)$. This behavior of the SFF in plateau period is more complex. It has been suggested that this can be reproduced by summing over higher order terms that appear in the genus expansion \cite{plateau1,SFF-APP-5}.\\
\\
$\blacksquare$ The fourth term: $\rho_0(E)\delta(E-E')$

By substituting it into (\ref{A-1}) we obtain $Z_D(2\beta)$ which is the constant value of the plateau period :\\
\begin{align}
\langle Z(\beta+iT) Z(\beta-iT)\rangle \ \sim Z_{D}(2\beta) \hspace{20pt} (T \gtrsim e^{S_{0}}). \label{A-9} \\ \nonumber 
\end{align}
At very late time, the SFF converges to $Z_{D}(2\beta)$ instead of decaying forever, suggesting that information is not completely lost.

\begin{figure}[H]
\centering
\includegraphics[width=115mm]{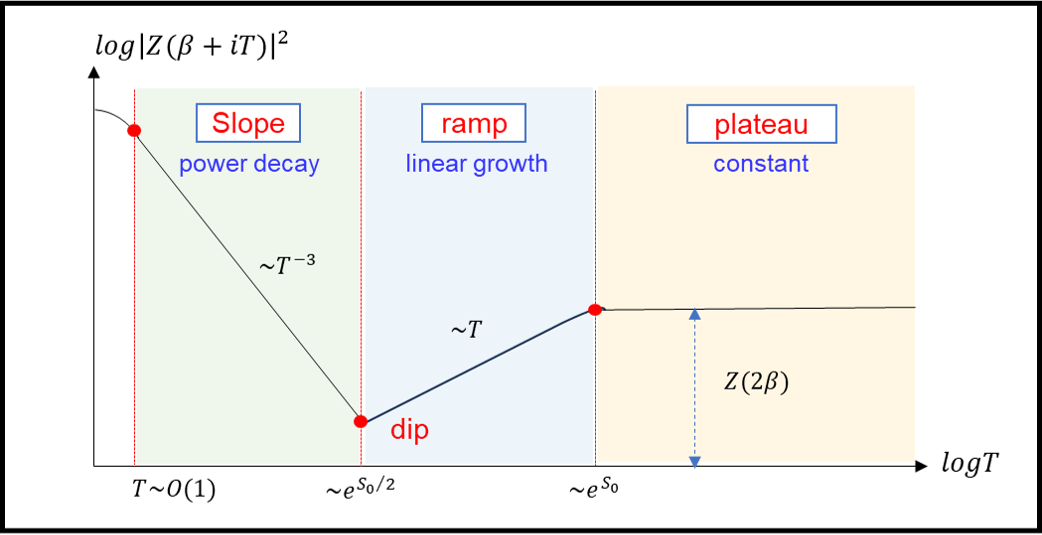}
\caption{\small Behavior of the SFF in JT gravity}
\label{figA-1(SFF)}
\end{figure}

The overall late times behavior of the SFF is summarized in Fig.\ref{figA-1(SFF)}.
\section{Derivation of (\ref{3-29}), (\ref{4-7}) and (\ref{4-4})}\label{Appendix-ERB-1}
Here we first explain the Hadamard regularization that is needed to obtain finite values for divergent integrals (\ref{3-29}) and (\ref{4-7}), and present some details of the derivation of (\ref{4-4}).

The Cauchy principal value and the Hadamard regularization are standard methods for obtaining finite values for divergent integrals. As an example, let's take \\
\begin{align}
\int^{\infty}_{-\infty} \frac{d\omega}{\omega^n} e^{-i\omega T}, \label{B-1} \\ \nonumber
\end{align}
with $n \geq 1$ and real $T$. The integral is divergent because of the pole at $\omega=0$. The Cauchy principal value for such integrals is defined by integrating over $D_{\epsilon}=:[-\infty,-\epsilon] \cup [\epsilon, \infty]$ and sending $\epsilon \to 0$. For $T \gtrless 0$ one finds the following results:\\
\begin{align}
&\int_{D_{\epsilon}} \frac{d\omega}{\omega} e^{-i\omega T} \ = \ \mp i\pi \label{B-2} \\
&\int_{D_{\epsilon}} \frac{d\omega}{\omega^2} e^{-i\omega T} \ = \ \mp \pi T + \frac{2}{\epsilon} \label{B-3} \\
&\int_{D_{\epsilon}} \frac{d\omega}{\omega^3} e^{-i\omega T} \ = \ \frac{\pm i\pi}{2}T^2 -\frac{2iT}{\epsilon} \label{B-4} \\
&\int_{D_{\epsilon}} \frac{d\omega}{\omega^4} e^{-i\omega T} \ = \ \frac{\pm \pi}{6}T^3 -\frac{T^2}{\epsilon} + \frac{2}{3\epsilon^3}. \label{B-5} \\ \nonumber
\end{align}
The Hadamard regularization means removing all terms with negative powers of $\epsilon$ before taking the limit $\epsilon \to 0$. Thus we obtain (\ref{3-29}) and (\ref{4-7}).

Now we turn to the derivation of (\ref{4-4}). By substituting $E_1=E+\frac{\omega}{2}, E_2=E-\frac{\omega}{2}$ into $|\mathcal{O}_{E_1,E_2}|^2$ defined in (\ref{2-5}) one obtains\\
\begin{align}
|\mathcal{O}_{E_1,E_2}|^{2} = \frac{|\Gamma(\Delta+2i\sqrt{2E})|^2|\Gamma(\Delta+1+i\frac{\omega}{\sqrt{2E}})|^2} {2^{2\Delta+1}\Gamma(2\Delta+1)}\cdot \frac{2\Delta}{\left( \Delta + \frac{i \omega}{\sqrt{2E}} \right)\left( \Delta - \frac{i \omega}{\sqrt{2E}} \right)}. \label{B-6} \\ \nonumber
\end{align}
In the limit $\Delta \to 0$ the two factors in the RHS become\\
\begin{align}
&\lim_{\Delta \to 0} \frac{|\Gamma(\Delta+2i\sqrt{2E})|^2|\Gamma(\Delta+1+i\frac{\omega}{\sqrt{2E}})|^2} {2^{2\Delta+1}\Gamma(2\Delta+1)} = \frac{\omega}{32\pi^2 E\hat{\rho}_0(E)\hat{\rho}_0(\omega)} \label{B-7} \\
&\lim_{\Delta \to 0} \frac{2\Delta}{\left( \Delta + \frac{i \omega}{\sqrt{2E}} \right)\left( \Delta - \frac{i \omega}{\sqrt{2E}} \right)} = 2\pi \sqrt{2E}\delta(\omega), \label{B-8} \\ \nonumber
\end{align}
where $\hat{\rho}_0(E)=\frac{1}{2\pi^2}\sinh(2\pi\sqrt{2E})$ and $\hat{\rho}_0(\omega)=\frac{1}{2\pi^2}\sinh\frac{\pi\omega}{\sqrt{2E}}$. By differentiating with respect to $\Delta$ and sending $\Delta \to 0$ one finds \\
\begin{align}
\frac{\partial}{\partial \Delta}|\mathcal{O}_{E_1,E_2}|^2 \ \xrightarrow{\Delta \to 0}& \  \frac{\delta(\omega)}{4\hat{\rho}_0(E)} \big\{ \psi(2i\sqrt{2E}) + \psi(-2i\sqrt{2E}) -2\log2 \big\} \nonumber \\
& + \frac{\omega}{8\pi^2 \hat{\rho}_0(E)\hat{\rho}_0(\omega)}\Big\{ \frac{1}{\omega^2+\epsilon^2} - \frac{2\epsilon^2}{(\omega^2+\epsilon^2)^2} \Big\}. \label{B-9} \\ \nonumber
\end{align}
Here we kept $\epsilon = \sqrt{2E}\Delta$ small but nonzero in the second term. We are going  to multiply (\ref{B-9}) by some function of $\omega$ which is regular at $\omega=0$ and then integrate over $\omega$. The small constant $\epsilon$ then serves as a regulation. We need to compare this with the Hadamard regularization.

Let $f(\omega) = \sum_{k \geq 0} f_k \omega^k$ be a function which is regular at $\omega=0$, and consider the integral:\\
\begin{align}
\int^b_a \frac{d\omega}{\omega^2} f(\omega) \quad (a<0<b) \label{B-10}. \\ \nonumber
\end{align} 
If one uses the Cauchy principal value, one obtains\\
\begin{align}
\int^{-\epsilon}_{a} \frac{d\omega}{\omega^2} f(\omega) + \int^{b}_{\epsilon} \frac{d\omega}{\omega^2} f(\omega) = f_0\Big( \frac{2}{\epsilon} + \frac{1}{a}-\frac{1}{b}\Big) +f_1 \text{ln} \left|\frac{b}{a}\right| + \sum_{k \geq 2} \frac{f_k}{k-1}\big( b^{k-1}-a^{k-1} \big). \label{B-11}  \\ \nonumber
\end{align}
On the other hand, by regularizing as in (\ref{B-4}) one obtains a finite result:\\
\begin{align}
\int^b_a \Big\{ \ \frac{1}{\omega^2+\epsilon^2} - \frac{2\epsilon^2}{(\omega^2 + \epsilon^2)^2} \ \Big\}f(\omega) d\omega \ \xrightarrow{\epsilon \to 0} \  f_0\Big( \frac{1}{a}-\frac{1}{b}\Big) +f_1 \text{ln} \left|\frac{b}{a}\right| + \sum_{k \geq 2} \frac{f_k}{k-1}\big( b^{k-1}-a^{k-1} \big). \label{B-12}  \\ \nonumber
\end{align}
Therefore, the regularization of (\ref{B-9}) by $\epsilon$ precisely corresponds to the Hadamard regularization of:\\
\begin{align}
\frac{\partial}{\partial \Delta}|\mathcal{O}_{E_1,E_2}|^2 &= \frac{\delta(\omega)}{4\hat{\rho}_0(E)} \big\{ \psi(2i\sqrt{2E}) + \psi(-2i\sqrt{2E}) -2\log2 \big\} + \frac{1}{8\pi^2 \hat{\rho}_0(E)\hat{\rho}_0(\omega)\omega}. \label{B-13} \\ \nonumber
\end{align}

\end{document}